\documentclass{sig-alternate}

\usepackage{graphicx}
\usepackage{booktabs}
\usepackage{array}
\usepackage{subfigure}
\usepackage{epsfig}
\usepackage{algorithm}
\usepackage{algpseudocode}
\usepackage{amsmath}
\usepackage{indentfirst}
\usepackage{amsmath,amssymb}
\usepackage{multirow}
\usepackage{url}
\usepackage{color, colortbl}

\newtheorem{definition}{Definition}
\newtheorem{pdef}{Problem Definition}

\newtheorem{example}{Example}
\newtheorem{prop}{{Proposition}}
\newtheorem{lemma}{{Lemma}}
\newcommand{\nop}[1]{}
\newcommand{\remind}[1]{{\small\color{red}{\bf hh: #1}}}

\begin{document}

\small\title{The Links Have It: Infobox Generation by Summarization over Linked Entities}
\numberofauthors{5}
\author{Kezun Zhang$^{\S}$,\; Yanghua Xiao$^{\S}$,\; Hanghang Tong$^{\ddag}$,\; Haixun Wang$^{\dag}$,\; Wei Wang$^{\S}$,\;\\
$^{\S}$\affaddr{\{kzhang12, shawyh, weiwang1\}@fudan.edu.cn}\;
$^{\ddag}$\affaddr{tong@cs.ccny.cuny.edu};
$^{\dag}$\affaddr{haixun@google.com}  \\
$^{\dag}$\affaddr{School of Computer Science, Shanghai Key Laboratory of Data Science, Fudan University, Shanghai, China}\\
$^{\ddag}$\affaddr{City College, CUNY, NY, USA}\\
$^{\dag}$\affaddr{Google Research, USA}
}

\maketitle

\begin{abstract}
\small
Online encyclopedia such as Wikipedia has become one of the best sources of knowledge. Much effort has been devoted to expanding and enriching the structured data by {\it automatic} information extraction from unstructured text in Wikipedia. Although remarkable
progresses have been made, their effectiveness and efficiency is still limited as they try to tackle an extremely difficult natural language understanding problems and heavily relies on supervised learning approaches which require large amount effort to label the training data.

In this paper, instead of performing information extraction over unstructured natural language text directly, we focus on a rich set of semi-structured data in Wikipedia articles: {\em linked entities}. The idea of this paper is the following: If we can summarize the relationship between the entity and its linked entities, we immediately harvest some of the most important information about the entity.
To this end, we propose a novel rank aggregation approach to remove noise, an effective
clustering and labeling algorithm to extract knowledge. 
We conduct extensive experiments to demonstrate the effectiveness and efficiency of the proposed solutions. Ultimately, we enrich Wikipedia with 10 million new facts by our approach.

\end{abstract}


\section{Introduction}

Online encyclopedia has become one of the best sources of knowledge. A typical
example is Wikipedia\footnote{\url{http://www.wikipedia.org}}, which contains 3.04 million articles for English language and covers a
wide range of human knowledge.  Another fast growing online
encyclopedia is BaiduBaike\footnote{\url{http://www.baike.baidu.com/}}, which contains 5 million entities and is
the largest knowledge base in Chinese. Wikipedia and
BaiduBaike are organized in similar ways, and have become the caliber of other online encyclopedias. In this paper, we focus on
these two encyclopedias for information extraction.

\begin{figure}[!htb]
\centering \epsfig{file=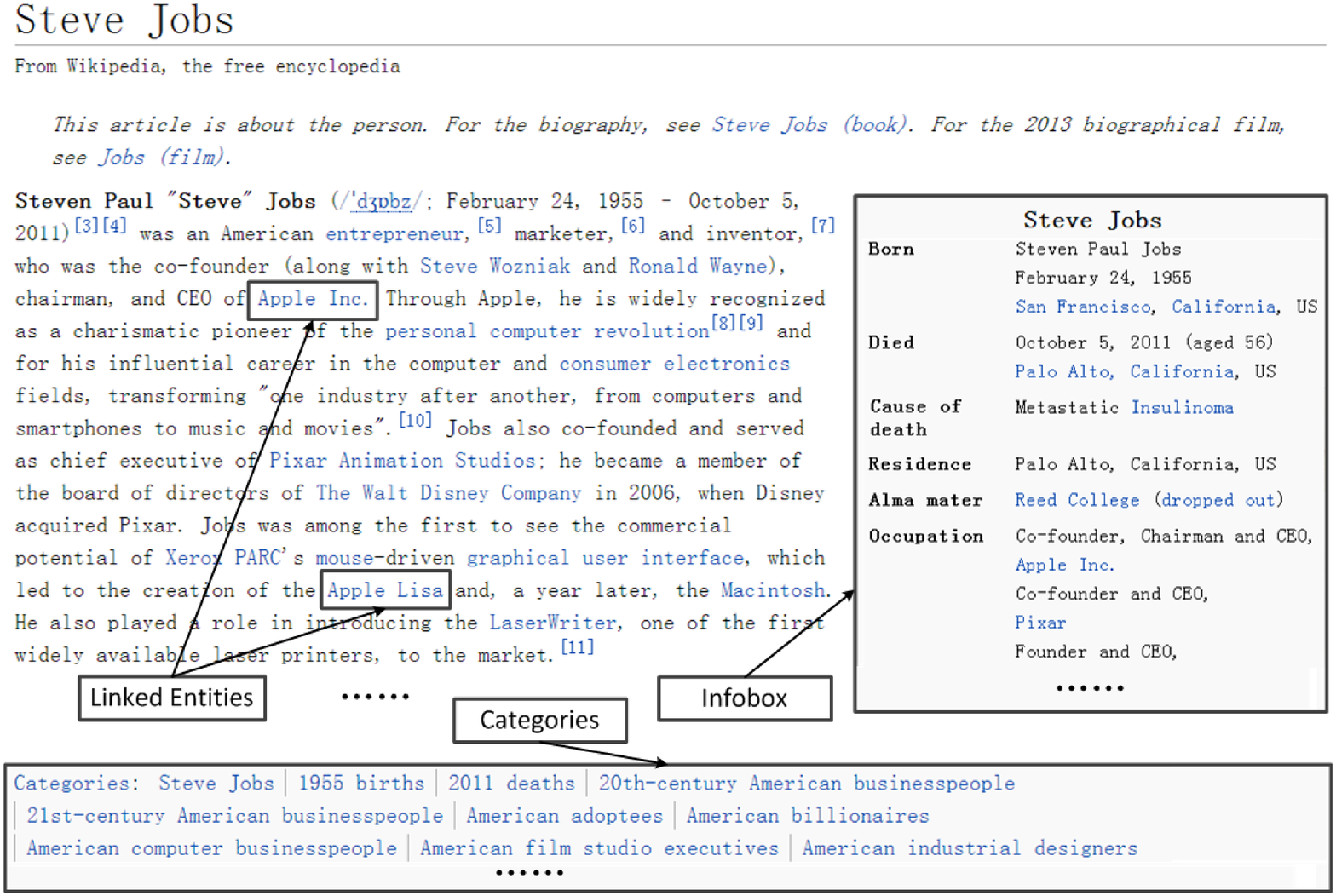,width=3.5in}
\caption{\small{Fragment of {\it Steve Jobs} in Wikipedia.}}\label{fig:stevejobs_wikipedia}
\end{figure}

Among others, one critical reason that makes online encyclopedias extremely valuable is that part of their data is structured, and hence
machine processible. Usually, a Wikipedia article is about an
entity. Many Wikipedia articles contain structured information such as
{\em table}, {\em image}, {\em text}, {\em citation}, etc., all of which are the targets of
information extraction. More importantly, many entities are associated
with an {\em infobox} which consists of a set of ($property$, $value$) pairs
about the entities.  As an example,
Figure~\ref{fig:stevejobs_wikipedia} shows the Wikipedia article about
{\it Steve Jobs}, with an infobox on the right side, wherein the
first property is {\it Born} and its value is {\it Steven Paul Jobs, February 24, 1955, San Francisco, California, US.}.  Such structured information is the core building block behind many
applications, including search engines, for answering user questions
about these entities, etc.


Despite much effort to enrich structured data, the current infobox in Wikipedia is often {\it incomplete} and {\it inconsistent}. This is mainly due to the fact that most infobox is generated by human editing, which is not just labor intensive but also error prone. To be specific,

\begin{itemize}
\item About 55\% Wikipedia articles do not have infobox. These are not
  only those less popular articles, but also new
  articles~\cite{InfoboxSuggestion}. For articles that have infobox,
  the information in the infobox is often incomplete. Some important
  properties may be missing, and values of certain properties may be
  incomplete~\cite{IBminer}.
\item Information in infoboxs is often inconsistent across different articles
  and entities. For example, the property {\it ''place of birth''} in
  some infoboxes is also expressed as {\it ''birthplace''} in other
  infoboxes; The property value {\it ''America''} and {\it
    ''United States (US), America''} refer to the same country, etc.
\end{itemize}

In order to address the drawbacks of human editing, recently, extensive effort has focused on expanding and
enriching the structured data by {\it automatic} information
extraction from unstructured text in Wikipedia~\cite{KnowItAll,Textrunner,IBminer}.  Although remarkable
progresses have been made, their effectiveness and scalability are still somewhat limited mainly for the two reasons. First, these methods rely on several natural
language understanding tasks (e.g., named entity recognition, dependency parsing, and relationship extraction), which themselves are extremely challenging and error
prone. Second, many of the existing approaches are costly, since they are essentially supervised learning methods, and hence require large amount of labeled training examples.


In this paper, we propose an alternative approach for enriching
structured data. Instead of performing information extraction over
unstructured natural language text directly, we focus on a rich set of
semi-structured data in Wikipedia articles: {\it linked entities}. A Wikipedia article typically consists of many links to other Wikipedia
articles. Intuitively, the author of the article, in describing a
Wikipedia entity, refers the reader to many other entities that are
important or related to the entity. The key idea of this paper is the
following: If we can summarize the relationship
between the entity and its linked entities, then we immediately
harvest some of the most important information about the entity.



\begin{minipage}{\textwidth}
\begin{minipage}{0.17\textwidth}
                \tiny
                \makeatletter\def\@captype{table}\makeatother\caption{\tiny{Entities}}
                \begin{tabular}[h]{|l|}
                \hline
                Toy Story\\ Cars (film)\\ Brave (2012 film)\\
                Intel\\ Dell\\ Apple Inc.\\
                \bf{\underline{blood pressure}}\\ \bf{\underline{The Public Theater}}\\
                Apple I \\ Apple Lisa\\
                Maria Shriver\\ Lev Grossman\\ \hline
                \end{tabular}
                \label{tab:stevejobs_linkedentities}
        \end{minipage}
\begin{minipage}{0.17\textwidth}
                \tiny
                \makeatletter\def\@captype{table}\makeatother\caption{\tiny{Knowledge}}
                \begin{tabular}[h]{ |p{2.3cm}|l| }
                \hline
                    \textbf{property} & \textbf{value} \\ \hline
                    \multirow{3}{*}{Pixar Animated Films} & Toy Story \\
                     & Cars (film) \\
                     & Brave (2012 film) \\ \hline
                    \multirow{3}{*}{Electronic Companies} & Dell \\
                     & Intel \\
                     & Apple Inc. \\ \hline
                    \multirow{2}{*}{American Writers} & Maria Shriver \\
                     & Lev Grossman \\ \hline
                    \multirow{3}{*}{Apple Inc. Hardware} & Apple I \\
                     & Apple Lisa \\ & \\ \hline
                \end{tabular}
                \label{tab:ExtractedKnowledge}
        \end{minipage}
\end{minipage}

Let us use the example in Figure~\ref{fig:stevejobs_wikipedia} to illustrate the intuition of our approach.
Table~\ref{tab:stevejobs_linkedentities} lists some linked entities in
the Wikipedia article of {\it Steve Jobs}, which cover a variety of different aspects of the article entity {\em Steve Jobs}. If we further convert these linked entities into something shown in
Table~\ref{tab:ExtractedKnowledge}, where we assign a property label to a
linked entity or a set of linked entities, the result provides a comprehensive, structured summary of the entity {\em Steve Jobs}.

In order to fulfill this basic idea, there are the following challenges we need to address as follows.

\paragraph*{C1. How to accurately summarize linked entities} In order to convert the unstructured linked entities list in Table~\ref{tab:stevejobs_linkedentities} to structured {\em (property, value)} pairs in Table~\ref{tab:ExtractedKnowledge}, we need to group the similar linked entities (i.e., values) together as well as assign a label (i.e., property) for each group. Here, our key observation is that
it is relatively easier to summarize a group of entities than an
individual because the group members disambiguate each other. We
thus propose a ''cluster-then-label'' approach: We divide linked
entities into different semantic groups, and then give each group a
semantic label (a property). More specifically, we propose a G-means
based clustering algorithm to cluster the linked entities into
different semantic groups. In the {\it Steve Jobs} example, we obtain four
clusters. We further propose a label generating algorithm to generate
a label for each group. Each labeled group is eventually a candidate
($property$, $value$) pair for the infobox.

\paragraph*{C2. How to remove unrelated linked entities} 
Although most linked entities are semantically related to the
article entity, some might have weak or no semantic relevance
to the article entity. 
Take the {\em Steve Jobs} example again,  we can see that some linked entities (e.g., {\it blood pressure} and
{\it The Public Theater}, etc) are not related to {\em Steve Jobs}. To remove these irrelevant linked entities, we propose a novel ranking aggregation approach
that integrates different ranking mechanisms to detect noisy linked entities.

\paragraph*{Contributions}
In summary, this paper proposes an alternative, radically different approach for infobox generalization for online encyclopedia. By focusing on linked entities, we bypass all the difficulties posed by the existing approaches, including the challenging NLP tasks, manual labeling and human editing. More specially, the main contributions of the paper are three-fold. First, to extract knowledge from the linked entities, we propose an effective clustering and labeling algorithm. Second, we propose a novel rank aggregation approach to detect and remove noisy linked entities for wiki articles.
Third, we conduct extensive experimental evaluations to show that our method generates comprehensive infobox with better quality.

The rest of the paper is organized as follows. In section 2, we give a detailed
description of handling noisy linked entities. In section 3, we
introduce the "cluster-and-label" algorithm.
Datasets and experiments are
described in section 4. In section 5, we introduce some related works. In section 6, we conclude our paper.


\section{Remove Noisy Linked Entities}
In this section, we show how we remove noisy linked entities. We first show that the noisy entities are nontrivial problem in online encyclopedias by empirical studies. Then, we propose a novel ranking aggregation approach to identify the noisy linked entities.

\nop{
\paragraph*{Problem Statement}
Given an article $a$ in online encyclopedia and its linked entities set $l(a)$, which contains the noisy linked entities and related ones. In this section, we use rank aggregation based ranking measure to rank entity in $l(a)$, then the low ranked entities are viewed as noises and be removed.
}

\subsection{Empirical Studies}

In typical online encyclopedias, some linked entities have weak relevance to the article entity. These entities become noises for the understanding of the semantic of the article entity. For example, {\it 'Steve Jobs'} in Wikipedia also has links to {\it blood pressure, The Public Theater} etc., each of which obviously has a weak relationship to {\it 'Steve Jobs'}. They are linked just because they have a corresponding entry in the knowledge base. We need to identify and remove them.

Next, we design an experiment to show that noisy entities are not trivial phenomenon. That is, most articles have noisy linked entities. For each article in Wikipedia or BaiduBaike, we calculate a semantic distance between the article and each of its linked entity. In our study, we use {\it Google Distance Inspired} distance~\cite{WLM}, which is defined as
\begin{equation}
sr(a,b)=\frac{log(max(|A|,|B|)-log(|A\bigcap B|)))}{(log(|W|)-log(min(|A|,|B|)))}
\end{equation}, where $A(or\ B)$ represents linked entities of article $a(or\ b)$, and $W$ represents entire articles in Wikipedia. We regard the linked entity as noise if the distance is larger than threshold 0.53.

\begin{figure}[h] \centering
\subfigure[\small{Wikipedia}] { \label{fig:wiki_noisy_link_distribution}
\includegraphics[width=0.45\columnwidth]{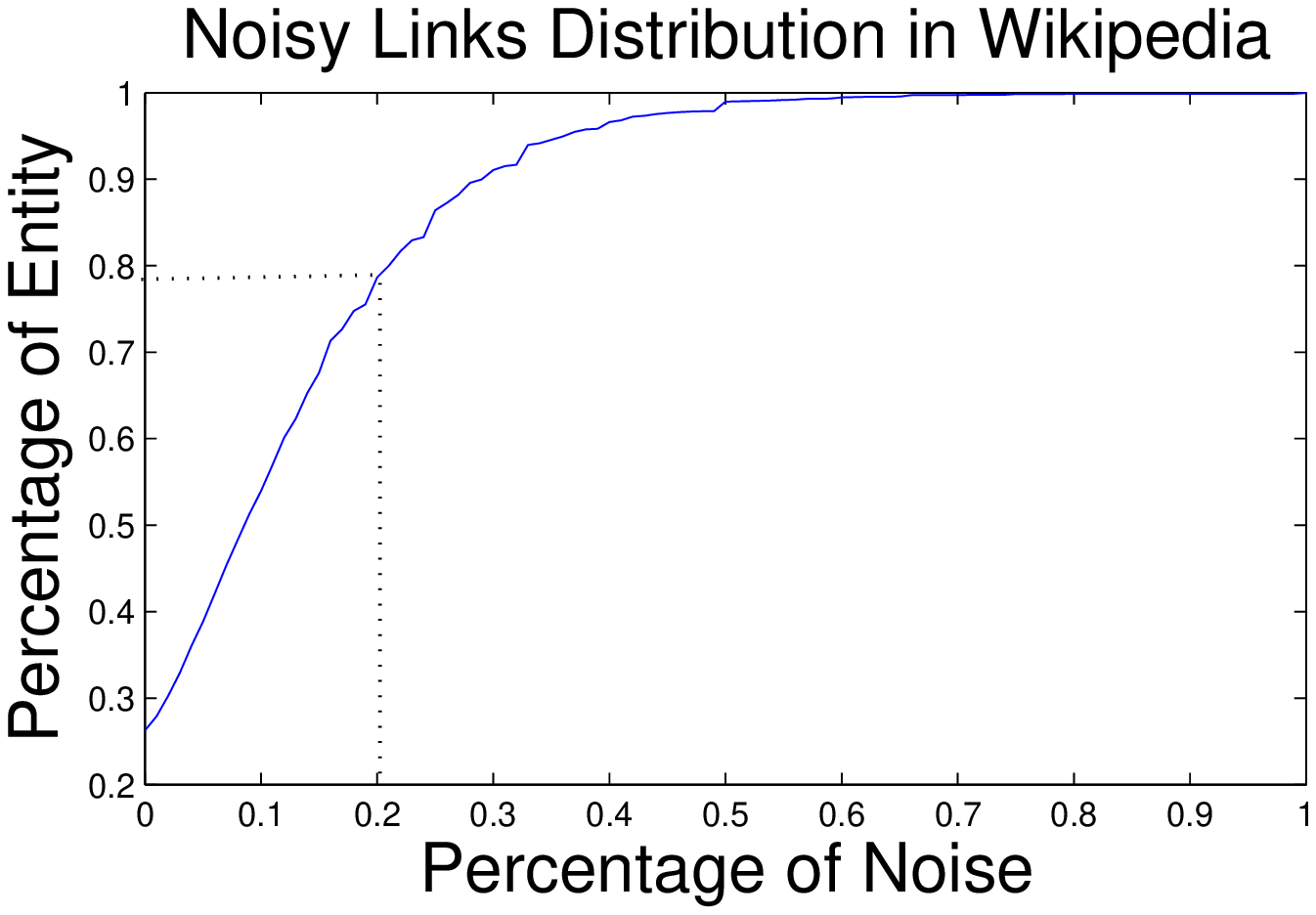}
}
\subfigure[\small{BaiduBaike}] { \label{fig:baidu_noisy_link_distribution}
\includegraphics[width=0.45\columnwidth]{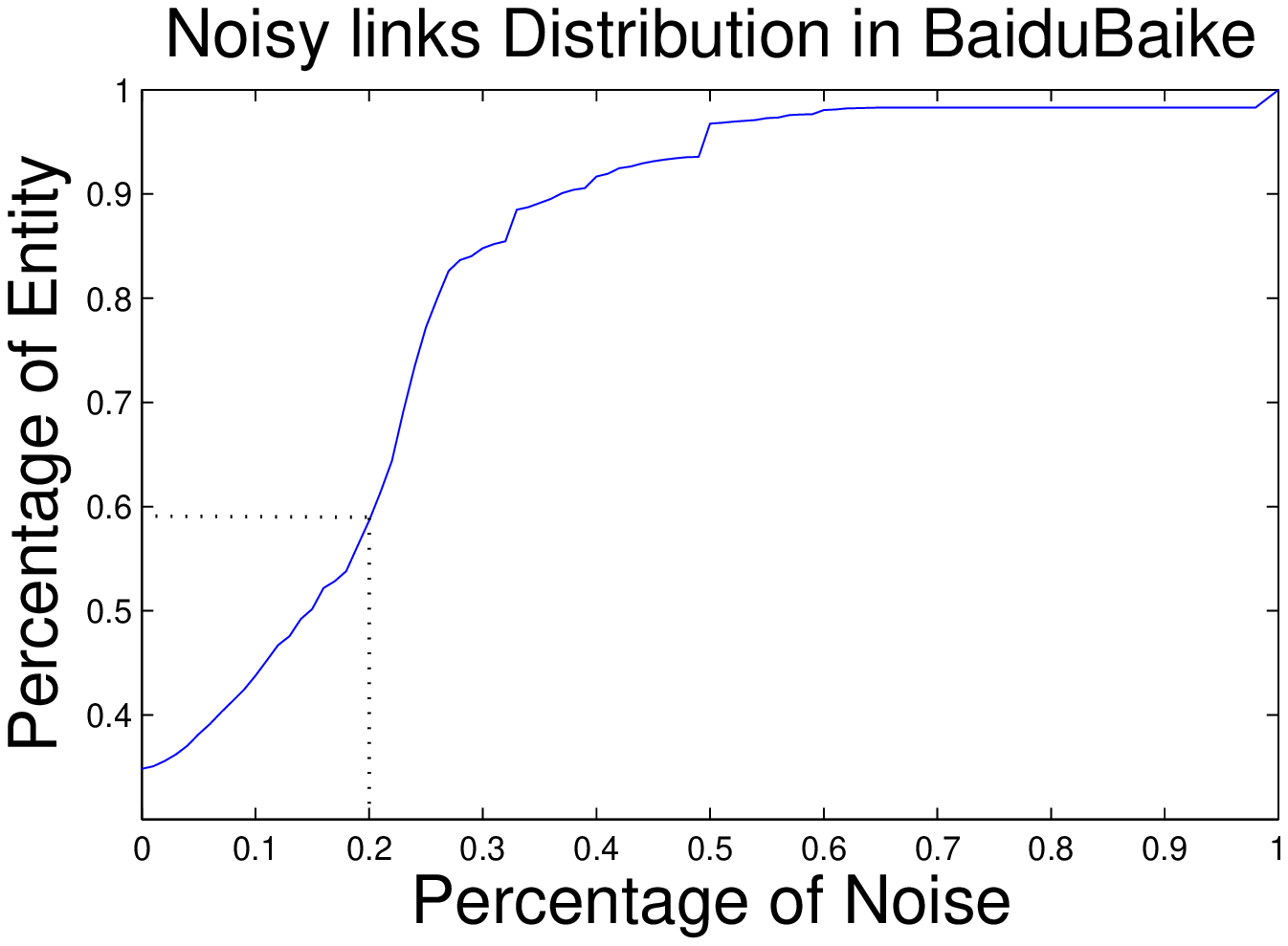}
}
\caption{\small{Noisy Linked Entities Distribution in Wikipedia and BaiduBaike.}}
\label{fig:link_noise}
\end{figure}

We summarize the cumulative distribution of the percentage of noisy links, and the results on Wikipedia and BaiduBaike are shown in Figure~\ref{fig:link_noise}. We found that in Wikipedia nearly 73\% of articles have noisy linked entities (only 27\% articles have no noisy entities), and 21\% articles have more than 20\% noisy linked entities. In BaiduBaike, nearly 80\% articles have noisy links (20\% articles have no noisy entities), and 42\% articles have more than 20\% noisy linked entities. The existence of the noisy linked entities makes it difficult to accurately understand the semantic understanding of entities. 


\subsection{Position-aware Ranking Aggregation}
The basic idea to remove noisy linked entities is to rank all linked entities by their semantic relatedness to the article entity and then remove the semantically unrelated entities. Thus, ranking the semantic relatedness becomes a key issue. There are many individual ranking schemes of semantic relatedness. However, in general each individual ranking can only characterize the a specific aspect of the semantic relatedness. Thus, an aggregated ranking is necessary for the accurate identification of non-related entities. Many existing ranking aggregation approaches have been proposed. However most of them assume the uniform quality distribution of the ranking. That is the ranking results have the same quality for any two elements in the ordering. However, we found that the individual ranking we used in this paper has a non-uniform quality distribution, which motivates us to propose a position-aware ranking aggregation approach.

\paragraph*{Preliminaries}
We first formalize the preliminary concepts.
A ranking $r_i$ can be considered as a linear ordering on the linked entities. This means that given the linked entities set $U$ with $N$ elements, $r_i$ is an one to one mapping from $U$ to ${1,2,...N}$. We always assume that elements of higher or topper rankings have smaller value.
 The quality function $q_r$
of the ranking $r$, is defined as a function $q: \{1..n\}\rightarrow [0,1]$.
$q_r(i)$ measures our belief on the fact that the $i$-th element under the ranking $r$ owns the $i$-th ranking position. Hence, $q_r$
is a function of the position of the ranking. Suppose we are given two rankings $r_1, r_2$ such that their quality function have opposite monotonicity. That is, $q_{r_1}(i)$ increases and $q_{r_2}(i)$ decreases with $i$. Thus, $r_1(e)$ (or $n-r_2(e)$) quantify the quality of $e$ ranking $r_1$ (or $r_2$). We refer to them as the {\it credit} of $e$ in the ranking. The smaller $r_1(e)$ (or $(n-r_2(e))$) is, the more credit that $e$ owns in $r_1$ (or $r_2$).

\subsubsection{Metrics}
Our aggregated ranking is developed upon two wildly used measures, {\it co-occurrence} based metric and {\it overlap coefficient}. This subsection elaborates these two measures.

\paragraph*{Co-occurrence}
Entities may co-occur in a common page as linked entities. If two entities always co-occur in a page, they are more likely semantically related. For example, {\it 'milk'} and {\it 'bread'} always co-occur in pages describing food and hence they are relevant in semantic.
We use PMI (pointwise mutual information) to measure the degree of the co-occurrence for a pair of entities. $PMI$ of entity $x$ and entity $y$ is defined as:
    \begin{equation}
    PMI(x,y)=log\frac{p(x,y)}{p(x)p(y)}
    \end{equation}, where $p(x,y)$ is the probability that $x$ and $y$ co-occur in the same entity page, $p(x)$ (or $p(y)$) is the probability that entity $x$ (or $y$) occurs in all co-occurrence pairs. $PMI$ is zero when $x$ and $y$ are independent, and maximizes when $x$ and $y$ are perfectly related (i.e., when $p(x,y)$ equals to $p(x)$ or $p(y)$). Compared to the direct co-occurrence number, $PMI$ evaluate their relatedness by statistical independence, which penalizes the independent pairs with high co-occurrence number.

\paragraph*{Overlap coefficient}
    Entities may share some common linked entities. A pair of entities has a larger overlap of linked entities is intuitively more relevant in semantic. For example, the closely-related entity pair {\it 'milk'} and {\it 'bread'} share a large number of common linked entities like food and drinks. We use the {\it Weighted Jaccard Coefficient }(WJC) to quantify the overlap ratio for an entity pair.  For two entities $x$ and $y$, $WJC$ is defined as:
    \begin{equation}
    \label{eq:wjc}
    WJC(x,y)=\frac{\sum_{e\in{N_x \cap N_y}}w(e)}{\sum_{e'\in{N_x \cup N_y}}w(e')}
    \end{equation}, where $N_x$ is the linked entities of $x$. Here, $w(e)$ is used as the weight of $e$, defined as:
     \begin{equation}
    \label{eq:idf1}
    idf(e)=log\frac{N-n(e)+0.5}{n(e)+0.5}
    \end{equation}, where $N$ is total number of articles, and $n(e)$ represents the number of articles containing a link to entity $e$. Compared to the naive $Jaccard$, $WJC$ use the $idf(e)$ as the weight to suppress the general entities. Like Jaccard coefficient, the higher the WJC is, the more related the entity pair is. If all entities have the same weight, $WJC$ will degrade into the naive Jaccard coefficient.

\paragraph*{Non-uniform Quality Distribution of Rankings}
We have two findings about these two rankings.
\begin{enumerate}
\item First, {\it the quality of an element under each ranking varies with its position in the the ranking}. That is to say, both $q_{WJC}(i)$ and $q_{PMI}(i)$ depends on $i$.

\item
Second, {\it $q_{WJC}(i)$ and $q_{PMI}(i)$  have opposite monotonicity}. In our case, we found that $PMI$ is good at identifying the noisy entities, but $WJC$ is good at discovering the strongly related entities. These findings imply that the we should develop position aware aggregation approaches.

\end{enumerate}

We give an example about entity {\it Apple Inc.} to justify the above two findings. More support will be found in the experiment sections.
We compare the ranked list of $WJC$ and $PMI$ as well as the aggregated measure that will be propped in the following text in Table~\ref{tab:ThreeLabeling} .
We can see that $WJC$ can recognize the strongly related entities and $PMI$ can correctly find the noisy linked entities. But $WJC$ regard related entities as noises (e.g. {\it iPhone 5} ) and $PMI$ regard the unrelated entity (e.g. {\it Software Update}) as related entity. In contrast, our ranking aggregation method take advantage of both two individual measures has less false positive and false negative results.

\begin{table}[h]\tiny
\caption{\small{Different ranking strategies for {\it Apple Inc.}}}
\label{tab:ThreeLabeling}
\begin{tabular}{ |p{2.5cm}|p{2.5cm}|p{2.5cm}| }
\hline
  \textbf{WJC} & \textbf{PMI} & \textbf{AGGREGATION}\\
  \hline
  Macintosh & Apple Battery Charger & Apple Worldwide Developers Conference\\
  Steve Jobs & Magic Mouse & OS X Mountain Lion\\
  OS X & Fortune (magazine) & Steve Jobs\\
  Apple Worldwide Developers Conference & Apple Inc. advertising & Apple TV\\
  OS X Mountain Lion & Software Update & MacBook Pro\\
  ... & ... &...\\
  Apple Time Capsule & Ireland & Cork (city)\\
  Business Model & Cork (city) & Video Calling\\
  Greenpeace International & Chancellor of the Exchequer & Broadway Books\\
  iPhone 5 & India & Greenpeace International\\
  \hline
\end{tabular}
\end{table}

\nop{
two kinds of state-of-art semantic relatedness measuring methods. One is corpus-based approaches, which measure semantic relatedness by analyzing large collected documents, e.g. WikiRelate~\cite{WikiRelate} and ESA~\cite{ESA}. But these methods are usually complicated and high cost~\cite{WLM}. Another is the lightweight measure such as {\it Google Distance Inspired} method~\cite{WLM}, which use linked entities only in Wikipedia. However, individual measure has its limitation in semantic representing, to handle this we propose an measures aggregated ranking mechanism.
}

\subsubsection{Position-aware Ranking Aggregation}
Our new ranking aggregation is based on the linear combination.
Given two rankings $r_1, r_2$ on $U$, the generic linear combination define the combined ranking $\sigma$ as
\begin{equation}
\label{eq:naive_aggregation}
\sigma(e)=\alpha\times r_1(e)+(1-\alpha)\times r_2(e)
\end{equation} for any $e\in U$, where $\alpha$ is used to control the preference to different rankings. In the naive linear combination, $\alpha$ is a static constant. That is, we use the same $\alpha$ for any $e\in U$.

However, previous observation implies that the preference to different rankings is dependent on the position of the entity under different rankings. Hence, in our new ranking aggregation we regard $\alpha$ as a function of $r_1(e)$ and $r_2(e)$ so that it can express the best preference to rankings for different entities.
Specifically, we define $\alpha(e)$ as:
\begin{equation}
\alpha(e)=\frac{1}{1+[\frac{r_1(e)}{(n-r_2(e))}]^{-\beta}}
\label{eq:alpha}
\end{equation} where $\beta$ is a parameter used to control the speed that the curve approaches to the climax.
Based on $\alpha(e)$,  we define our new scoring function of $e$
\begin{equation}
score(e)=\alpha(e)r_1(e)+(1-\alpha(e))r_2(e)
\end{equation}
When $\beta=1$, we have the new scoring function as
\begin{equation}
score(e)=\frac{nr_2(e)+ r_1^2(e)-r_2^2(e)}{n-r_2(e)+r_1(e)}
\label{eq:score}
\end{equation} It is easy to check that $score(e)\in [1, N]$.

Given the new score values of linked entities, we first normalize them. We use some articles as training data and label their linked entities as related or unrelated. We build a binary classification model and draw its ROC curve, finding that $0.77$ is the best threshold to distinguish unrelated linked entities from others. We use this threshold for all the other articles. 

\nop{
To compute the new ranking, in general we need set an appropriate value for $\alpha$.
Before we can derive a better aggregated ranking, we first study the influence of  $r_i(u)$ and $r_j(u)$ on $\sigma(u)$. As observed above, given an ordering $r_i$ and any two elements $u, v$, we may have different confidence on $r_i(u)$ and $r_i(v)$. For an ordering $r_i$, the confidence of $r_i(u)$ may monotonically increase/decrease with $r_i(u)$. For example, in our study, $PMI$ performs better in detecting noisy entities, and $WJC$ does well in finding strong related entities. That means when $PMI(u)$ is larger, $PMI$ is more plausible. In contrast, $WJC(u)$ is more plausible when $WJC(u)$ has a higher rank.
Hence, {\it the preference to $PMI(u)$ or $WJC(u)$ when defining $\sigma(u)$ depends on the value of $PMI(u)$ or $WJC(u)$.} In other words,......
}

\paragraph*{Rationality}
Next, we show how we derive the new ranking.
Given two rankings with oppositely monotonic quality functions, the aggregated ranking should bias towards to the one with higher quality. Specifically, for any entity $e$, we evaluate $\sigma(e)$ according to $\frac{r_1(e)}{n-r_2(e)}$. There are three specific cases:
\begin{enumerate}
\item Case 1: $r_1(e)/(n-r_2(e))\approx 1$. In this case, $e$ owns similar credit in $r_1$ and $r_2$. Hence, $r_1(e)$ and $r_1(e)$ should be assigned a similar weight close to 0.5.
\item Case 2: $r_1(e)/(n-r_2(e))> 1$.
 In this case, $e$ owns more credits in $r_1$ than in $r_2$. Hence, $\alpha$ should bias toward $r_1(e)$. That means the weight of $r_1(e)$ should be larger than 0.5 in the linear combination.
\item Case 3: $r_1(e)/(n-r_2(e))< 1$. It is the reverse case of Case 2. In this case, $\alpha$ should bias toward $r_2(e)$.
\end{enumerate}

\nop{
Hence, {\it the preference to $PMI(u)$ or $WJC(u)$ when defining $\sigma(u)$ depends on the value of $PMI(u)$ or $WJC(u)$.} {\bf In other words,.....}The start point of our method follows a simple idea, aggregation method should assign a high weight to $PMI$ if is judged a higher position, which increase the probability to be noise. On the contrary, aggregation should assign a height weight to $WJC$ if is preferred a lower position, which increase the probability to be strong related entity. Similarly, aggregation should balance the weight of $PMI$ and $WJC$ according to the ranking position to justify the probability to be noise. In the following, we will introduce our rank aggregation method at length.
}

\begin{figure} \centering
\subfigure[\small{Ratio of entities}] { \label{fig:ratio_a}
\includegraphics[width=0.45\columnwidth]{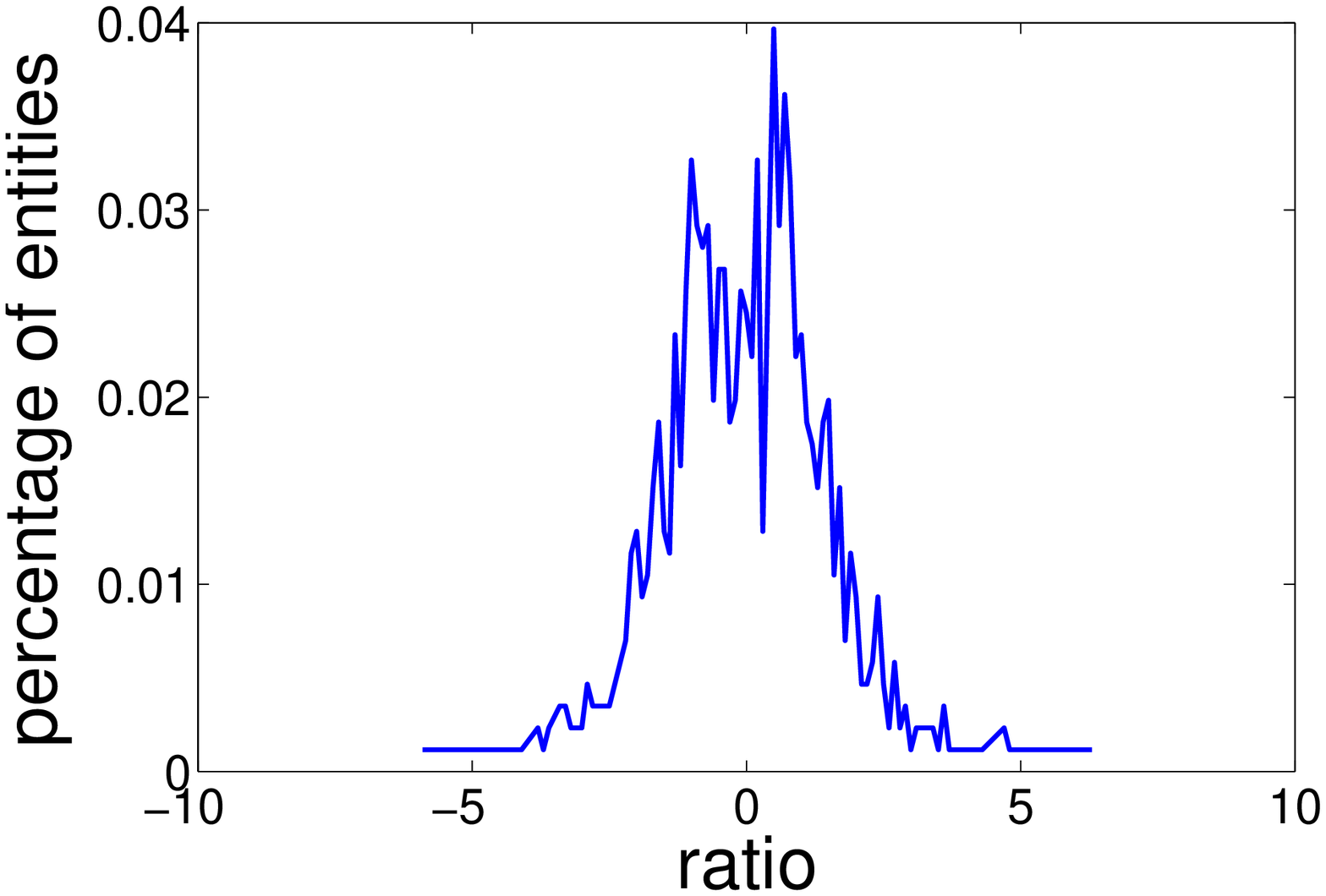}
}
\subfigure[\small{$\alpha$ as a function of $ratio$}] { \label{fig:ratio_b}
\includegraphics[width=0.45\columnwidth]{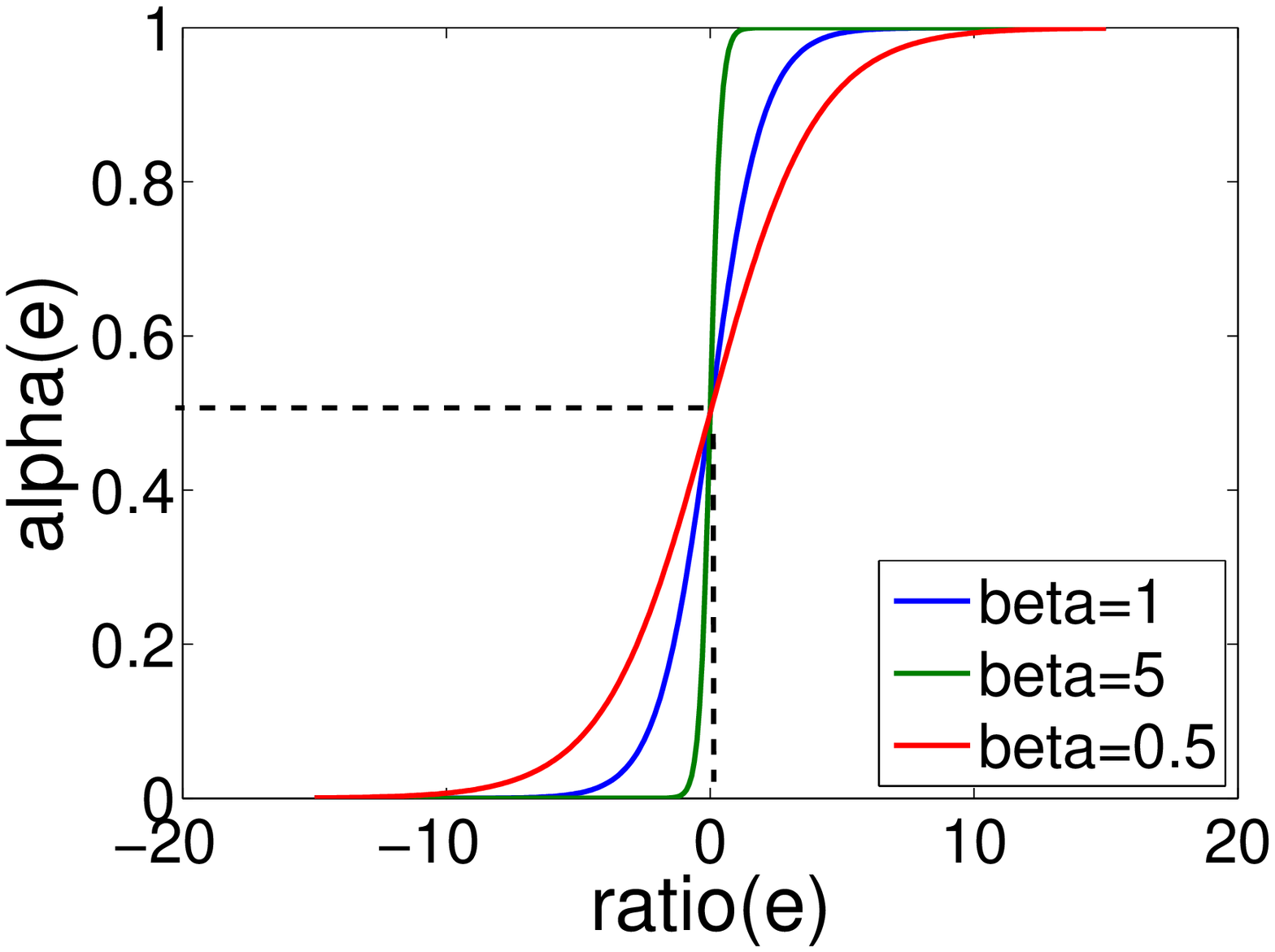}
}
\caption{Ratio and alpha.}
\label{fig:rank_ratio}
\end{figure}

Clearly, a {\it sigmoid} function can express the desired relationship between $\alpha$ and the ratio. Specifically, we use the most widely used {\it logistic} function $s(x)$, which is defined as:
$$s(x)=\frac{1}{1+exp(-\beta x)}$$, where $\beta$ is a parameter used to control the speed that the curve approaches to the climax.
Furthermore, if we replace the ratio by the its log ratio, all the requirement in the three cases can be satisfied. The log ratio is defined as:
\begin{equation}
\label{eq:aggr_ratio}
ratio(e)=ln\frac{r_1(e)}{n-r_2(e)}
\end{equation}
Substituting $x$ with the  the log ratio, we have the $\alpha$ as defined in Eq.~\ref{eq:alpha}.

\paragraph*{Selection of $\beta$}
We use 3380 linked entities of {\it Shanghai, Apple Inc., Steve Jobs, China, New York City, Barack Obama} as samples. For each of these linked entity, we calculate its $ratio$ function value (we use PMI and WJC as $r_1$ and $r_2$, respectively). We plot their distribution of $ratio$ in Figure~\ref{fig:ratio_a}. From the distribution, we can see that most $ratio$ values lie in the range of [-3,3], which hosts 95\% of all sampled linked entities. We also give the simulation of $\alpha$ as a function of $ratio(e)$ (see Eq.~\ref{eq:alpha}) with $\beta$ set as different values in Figure~\ref{fig:ratio_b}. The simulation shows that the larger $\beta$ is, the sharper increase happens around 0. The simulation also reveals that when $\beta=1$ the range of $ratio$ in which a significant $\alpha$ can be derived almost overlaps with the real range observed from the samples. Hence, typically we set $\beta=1$.



\section{Clustering and Labeling}
After removing the noisy linked entities, we keep only the semantically related linked entities. Next, we use a G-means based clustering approach to divide them into different semantic groups. Then, we label each group with an appropriate property name. In this way, we discover a new property and its value for an entity from its linked entities. Distance metric is key for a clustering algorithm. Hence, we first elaborate the distance metric.

\nop{
\paragraph*{Problem Statement}
Given a set of semantic related linked entities $l(a)$ for article $a$, we use cluster-then-label method to produce knowledge, and outputs clusters and corresponding labels for $a$. In this way, entities in a cluster and corresponding label make up a group of knowledge, which label stands for the property and the values for the property are the entities in the group. In this section, we first use Gmeans clustering approach to group entities in $l(a)$, then we use LCA(least common ancestor) based approach to label each group.
}

\subsection{Feature Selection and Distance Metric}

To define the distance metric, we first need to identify the effective features to characterize the objects to be clustered. Here, we use category information of entities for the clustering. In Wikipedia or BaiduBaike, an entity is usually assigned one or more categories by editors. A category is widely used to represent the concept of an entity. Hence, if a pair of entities has the similar categories, they probably belong to same concept (or domain, topic). Category or concepts information has been shown to be effective for the document clustering~\cite{CatDoClustering} and topic identification~\cite{CatNetwork}, which motivates us to use categories to construct the feature vector for the entity.

\paragraph*{Problem statement}
The naive solution is using direct categories of entities as the features. Let $F_e$ be the feature set for entity $e$. In the naive solution, $F_e$ contains all the direct categories of $e$. Let $n$ be the number of all categories in Wiki. We define a $n$-dimensional feature vector for each entity $e$, i.e., ${f_e}=<w(c_1), \ldots ,w_(c_{n})>$, where $w(c_i)$ measures {\it the significance that concept $c_i$ characterizes $e$}.
In general, $w(c_i)=0$ when $c_i$ is not in $F_e$, otherwise, $w(c_i)$ is defined by a certain measurement (such as {\it tf-idf} functions, we will elaborate it in later texts). Given two feature vectors of two entities $a, b$, their distance is defined by the cosine distance:
\begin{equation}D(a,b)=1-\frac{f_a\cdot f_b}{\|{f_a}\|\cdot \|{f_b}\|}
\label{eq:dist}
\end{equation}

\begin{figure}[!htb]
\centering \epsfig{file=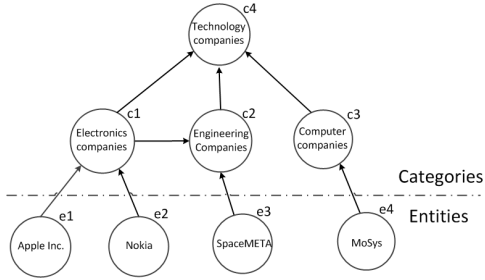,width=2.5in}
\caption{Category graph in Wikipedia.}\label{fig:cate_graph}
\end{figure}

However, using the direct categories for the distance metrics has the following two weaknesses:
\begin{itemize}
\item First, many categories are not hypernyms of the entity. Some categories express the semantics other than IsA relationship. For example, {\it Steve Jobs} has category {\it American Buddhists}, which is an IsA relationship. But it also has {\it 1955 births} (a property), {\it Apple Inc} (works-for relationship) and many other semantics other than IsA. In general, it is hard to use these non-IsA categories to characterize the concept of an entity.
\item Second, many direct categories are quite specific. We calculate the frequency of all categories in Wikipedia. We found that among the top-100 most frequent categories, 75\% is in the form of '? year of birth' or '? year of death'. Obvious these are specific categories that characterize a specific property of the entity. In general, the more specific the category is, the less possible two semantically-close concepts can be matched in terms of the category. For example, in category graph, shows in Figure~\ref{fig:cate_graph}, {\it Apple Inc.} can only match with {\it MoSys} (an IP-rich fabless semiconductor company) in term of a more abstract category {\it technology companies} instead of the specific one ({\it Electronics companies}). \end{itemize}

\begin{algorithm}[h]
\caption{Feature Selection and Weighting Algorithm}
\label{alg:feature}
\small
\begin{algorithmic}[1]
\Require {Entity $e$, Set of concept-category pair $C$}
\Ensure {Feature and corresponding weight of $e$}
\State IsA taxonomy graph $G\gets$ {\bf IsA-Construction($C$)};
\State $A_e\gets$ reachable categories from $e$ in $G$;
\For {$c$ in $A_e$}
    \State weight of $c$:$w_c=p(c|e)*idf(c)$, as in Eq.~\ref{eq:word_weight} to Eq.~\ref{eq:wf_idf};
    \State mark $c$ as a feature of $e$, corresponding weight is $w_c$;
\EndFor
\State \Return
\\
\Function{IsA-Construction}{$C$}
    \State $G=\phi$: IsA taxonomy graph, a directed graph;
    \State $\alpha$: a threshold parameter;
    \For {each $concept,category$ in $C$}
        \State weight of $category$ for $concept$ is calculated by Eq.~\ref{eq:cate_weight};
        \State add an edge <$concept$,$category$> to $G$ if weight > $\alpha$;
    \EndFor
    \State \Return $G$;
\EndFunction
\end{algorithmic}
\end{algorithm}

\paragraph*{IsA taxonomy construction}
To overcome the above weaknesses, we need to extend the feature set from the direct categories to high level categories, described in Algorithm~\ref{alg:feature}. We may recursively use the categories of the categories for the expansion. However the extension is not trivial. Because we need to ensure the expanded category can characterize the entity accurately. That is to say we expect to improve the recall without sacrificing the precision. For this purpose, we generally need a certain constraints on the extension to ensure the accuracy.
A general constraint is to only select the categories that are hypernyms of the entity. Because a hypernym is a concept of the entity, which is a natural interpretation of the entity. Thus, the problem is reduced to identification of a category that is a hypernym of an entity. We define a scoring function $p(c|e)$ to characterize the confidence on category $c$ being a hypernym of entity $e$.

The definition of $p(c|e)$ depends on the hierarchal structure of the hypernyms of $e$. For each entity $e$, we can {\it construct a high-quality hierarchical taxonomy just according to the Wiki categories}. The taxonomy for entity $e$, denoted by $G_e(V_e, E_e, w_e)$, is a {\it direct acyclic graph} with each edge $<c_1,c_2>$ assigned a weight $w(<c_1,c_2>)=p(c_2|c_1)$ which reflects our confidence on the fact that $c_2$ is a hypernym of $c_1$.

\paragraph*{\bf Algorithm to construct the taxonomy}
Given a threshold parameter $\alpha$, we construct the IsA taxonomy $G_e(V_e, E_e, w_e)$ for an entity or category $e$ by a level wise solution. Let $C=C^{0}=\{e\}$.
Suppose we have finished the $i$-th level ($i$ starts from 0). The $(i+1$-th level is as follows. For each category $c_2$ of any element (say $c_1$) in $C^{i}$ such that $p(c_2|c_1)\geq \alpha$, We add the direct edge from $c_1$ to $c_2$ into $E_e$ and use $p(c_2|c_1)$ as the edge weight. And add $c_2$ into $C$ and $V_e$ if $c_1 \notin \bigcup_{0\leq j\leq i}C^{j}$. These newly added categories constitute $C^{i+1}$. We add the direct edge from $c_1$ to $c_2$ into $E_e$ and use $p(c_2|c_1)$ as the edge weight. The procedure is repeated until no more valid category can be found.
It is easy to prove that {\it $G_e$ is a direct acrylic graph.}


\paragraph*{Scoring functions}
Next, we define $p(c|e)$. An observation is that many real hypernyms contains many frequent occurring words among the categories of the entities. This inspiration implies that we can use the word frequency to define $p(c|e)$. Specifically, for an entity or category $e$ and its categories $cat(e)$ in Wiki. We first score the words in hypernym $c$ for $e$. Let $f(s)$ be the number of categories in $cat(e)$ that contains word $s$. We have
\begin{equation}
\label{eq:word_weight}
p(s|e)=\frac{f(s)}{|cat(e)|}
\end{equation}
Let $k_c$ be the number of unique words in $c$. The confidence that the category $c$ is an appropriate hypernym of $e$ is defined as:
\begin{equation}
\label{eq:cate_weight}
p(c|e)=\frac{1}{k_c}\sum_{w\in c}{p(s|e)}
\end{equation}
Let $P_{ec}$ be the set of all the paths from $e$ to $c$ and $p_{ec}$ be one of such path. Now we are ready to define the confidence score for any category $c$ in $G_e$ as a hypernym for $e$.
\begin{equation}p(c|e)=\max_{p_{ec}\in P_{ec}}\prod_{<c_i,c_j>\in p_{ec}}p(c_j|c_i)
\end{equation}. The score is defined as the maximal accumulative product of the edge weight over all paths connecting $e$ to $c$. The larger the maximal produce, the more possible the concept is a hypernym of the entity. We give Example~\ref{exa:score} to illustrate our scoring functions.

\begin{example}[Scoring function]
Consider \emph{Apple Inc.}, its direct categories in Wikipedia are \emph{  \{electronics companies, home computer hardware companies, electronics companies of the united states, computer companies of the United States, steve jobs, apple inc., 1976 establishments in California, ...\} }. The most frequent words in the categories are \emph{\{companies, electronics, computer, united states\}}. Thus, the categories containing these words are likely hypernyms of \emph{apple inc.}, such as \emph{\{ electronics companies of the united states, electronics companies\}}. But \emph{steve jobs} will be dropped in our approach since it contains less frequent words. In the construction of the IsA taxonomy for the apple entity, some high-level categories such as \emph{technology companies} will be covered. Consequently, many indirect category will be used to characterize an entity.
\label{exa:score}
\end{example}

\paragraph*{Improved distance metric}
Finally, we are ready to define our improved distance metric, which share the same expression as Eq.~\ref{eq:dist} but with two improvements. First, $F_e$ is extend into $V_e-\{e\}$. That is all categories in $G_e$ except $e$ itself will be used as features.
Second, $w(c_i)$ is defined according to $p(c|e)$.
We use the {\it tf-idf} framework to define $w(c_i)$. We first define the $idf$ of a concept $c$, i.e., $idf(c)$ as
\begin{equation}
\label{eq:cate_idf}
idf(c)=log\frac{N}{|\{e|c\in V_e\}|}
\end{equation} where $N$ is the total number of entities in the Wiki and $|\{e|c\in V_e\}|$ is the number of entities whose IsA taxonomy contains $c$. Thus, the final weight of each feature is:
\begin{equation}
\label{eq:wf_idf}
w(c)=p(c|e)\cdot idf(c)
\end{equation}
To see the effectiveness of the above measurement, we rank the categories of entity {\it Apple Inc.} by $w(c)$ in Table~\ref{tab:concept_rank}. We can see that most categories of higher rank can characterize the entity accurately and expressively.

\begin{table}[!htb]
\caption{\small{Category ranking for {\it Apple Inc.}}}
\small
\begin{tabular}{|p{8cm}|}
\hline
computer companies of the united states\\ electronics companies\\ technology companies of the united states\\
networking hardware companies\\ retail companies of the united states\\ home computer hardware companies\\
...\\
steve jobs\\ apple inc.\\ warrants issued in hong kong stock exchange\\ \hline
\end{tabular}
\label{tab:concept_rank}
\end{table}

\subsection{Clustering Algorithm}
We may directly use K-means approach as the basic framework for clustering given the distance metric. But in our case, the naive K-means leads to bad results due to the following reasons.
\begin{enumerate}
\item First, in naive $K$-means the parameter K is specified by users, which is impossible when millions of entity clustering tasks need to be executed.
\item Second, the naive $K$-means randomly selection initial centers. The selection of initial center is influential on the final results \nop{{\bf citaiotn}}. A smart selection strategy is expected to obtain a better clustering result.
\end{enumerate}
To solve these problems, we propose a new clustering approach. The basic idea is using statistical test (proposed in G-means~\cite{g-means}) to guide the selection of best $K$, and using a dynamically center selection strategy (proposed in K-means++~\cite{kmeans++}) to determine the best initial central points.

Our clustering algorithm is described in Algorithm~\ref{alg:Gmeans}.
The algorithm accepts the set of data points $X$ as the input and return $K$ clusters. The algorithm recursively bi-partition the data until the stop criteria is reached.
The bi-partition procedure consists of three major steps:
\begin{enumerate}
\item In the first step, we select two data points $d_1, d_2\in G$ as the initial centers by $K$-means++~\cite{kmeans++}. $K$-means++ is smarter than the random generation of two cluster centers. It follows the principle that the probability of a datapoint to be center should be proportional to the distance from the already selected centers. Following the idea, we first choose a datapoint $d_1$ uniformly at random from the group $X$. Then, we select another datapoint $d_2$ from the group, with probability $$\frac{D(d_1,d_2)^2}{\sum_{x\in X}{D(d_1,d_2)^2}}$$, where $D(d_1,d_2)$ represents distance between $d_1$ and $d_2$.
\item In the second step, we run $K$-means on data points in $G$ with $K=2$ and the initial center as $d_1,d_2$. After the $K$-means reaches to the convergence state or gets maximal $iteration$, we get two clusters $G_1, G_2$ and their new centers $c_1,c_2$.
\item In the third step, project datapoint $d_i$ in $G$ onto vector $c_1-c_2$, which $d_i^{'}=d_i\cdot (c_1-c_2)/|(c_1-c_2)|$. And let $Z$ be the cumulative distribution of $d_i^{'}$. Finally, we test whether the Anderson-Darling statistic value $A_*^{2}(Z)$ lies in the range of non-critical values at significance level $\alpha$. If true, keep the original group and abandon the splitting. Otherwise, replace the group with two subclusters $G_1, G_2$ and continue bi-partition them until no new clusters emerging.
\end{enumerate}

\begin{algorithm}[h]
\caption{G-means Clustering Algorithm}
\label{alg:Gmeans}
\small
\begin{algorithmic}[1]
\Require {Datapoints $X$,significance level $\alpha$}
\Ensure {K Clusters}
\State $K\gets 1,G\gets X$
\State $Clusters\gets$ {\bf Bi-Partition(G,$\alpha$)}
\State \Return Clusters
\\
\Function{Bi-Partition}{$G,\alpha$}
    \State Select two datapints $d_1,d_2$ from group $G$ by $K$-means++;
    \State Run $K$-means with $k=2$ and the initial center as $d_1,d_2$;
    \State Let $G_1,G_2$ be the two clusters and $c_1,c_2$ be the corresponding two cluster centers;
    \If{$GaussianTest(G,c_1,c_2,\alpha)$} \Comment\textit{If datapoints in $G$ follow Gaussian distribution}
        \State \Return $G$
    \Else
        \State $K\gets K+1$
        \State \Return $Bi-Partition(G_1,\alpha)\cup Bi-Partition(G_2,\alpha)$
\EndIf
\EndFunction
\end{algorithmic}
\end{algorithm}

For example, we remove the noisy entities for {\it Apple Inc.} in Wikipedia, and cluster them in above algorithm, clusters show in Table~\ref{tab:apple_inc_cluster}.

\subsection{Labeling the Cluster}
Next, we assign a semantic label for each group. In this way, we explain why group of linked entities are linked to the article entity. The semantic label as well as the group of entities thus becomes a property of the target entity and its corresponding value. This information is a good supplement of the current infobox. For example, a cluster which contains \{ {\it Google Maps, ios 6, iBooks, xSan, iTunes}\}, If we assign the semantic label {\it ios software} for the cluster,  we successfully enrich the infobox of {\it Apple Inc.} with a property({\it ios software}).

That is the problem of cluster labeling, some researches have already conducted on cluster labeling. A popular method for labeling cluster is applying the statistic technologies to select frequency features. That is identifying the most common terms from the text that best represent the cluster topic. But the frequent terms may not convey meaningful message of the cluster. Because some popular terms are also frequently occur in other clusters.

As a result, an appropriate cluster label should characterize the common topic of entities in each cluster and simultaneously informative. A good cluster label should satisfy two requirements:
\begin{enumerate}
\item {\it Completeness}. It should cover most entities in the cluster. E.G. for first cluster in Table~\ref{tab:apple_inc_cluster}, label {\it tunisian-jewish descent} only cover one entity in the cluster. So we want a wildly covered label which can represents the group correctly.
\item {\it Informativeness}. We hope the label is the most specific label while covering all entities in the cluster. e.g. in first cluster {\it people by status} covers all entities in the cluster, but it is not informative.
\end{enumerate}
The completeness and the informativeness are contradicted to each other. In general, the more abstract a label, the more entities that it can cover. Some improvements have been done to generate a meaningful label. Inverse frequent term, takes both frequency and weight of a term into consideration. A meaningful label for a cluster is a term with maximal inverse frequency.

\paragraph{Baseline labeling strategies}
We first give two naive methods to label clusters. However, the naive solution in general has one or more weakness, which motivates us to a least common ancestor (LCA) based solution. In the previous subsection, we have built the IsA Taxonomy graph $G_e$ for each entity $e$. All categories in $G_e$ will be used for the labeling. Given a cluster $X=\{e_1,e_2,...,e_k\}$, let $\mathcal{C}$ be the union of each $V_{e_i}$. We have two baseline labeling strategies.

\begin{enumerate}
\item\textbf{Most Frequent Category} (MF for short).
The direct solution is labeling the cluster using the most popular category. Let $tf(c)$ be the number of $G_e$ such that $c\in V_e$ for all entities in the cluster. Thus, MF selection strategy is:
$$arg\max_{c\in \mathcal{C}}{tf(c)}$$
\item\textbf{Most Frequent yet Informative Category} (MFI for short)
Apparently, MF tend to select popular concept and most popular concept are abstract concept. Thus, the informativeness is sacrificed. To avoid this, we take the $idf$ like factor into account. Formally, MFI selection strategy is:
$$arg\max_{c\in \mathcal{C}}tf(c)\cdot idf(c)$$, $idf(c)$ is defined by Eq. ~\ref{eq:cate_idf}.
\end{enumerate}

However, the above labeling methods have the following weakness. 1) MF tends to select general (with good completeness) but less informative label. 2) MFI can recognize specific labels, but in many cases maybe over specific. Because some specific concepts own a large {\it idf} weight. Next, we propose a least common ancestor model to handle the tricky tradeoff between the informativeness and complexness.


\subsubsection{LCA based solution}
The LCA model is defined on the IsA Taxonomy graph for the cluster $X$ to be labeled.
Given a cluster $X=\{e_1,e_2,...,e_k\}$, we first construct the IsA Taxonomy graph for $X$, $\mathcal{G}_x$. We define $\mathcal{G}_x$ as the union of all IsA taxonomy graph $G_e$ such that $e\in X$. Here, we ignore the weight of $G_x$. Thus the {\it union} of two IsA taxonomy graphs $G_{e_1}$ and $G_{e_2}$ is the graph $G'(V', E')$ with $V'=V_{e_1}\cup V_{e_2}$ and $E'=E_{e_1}\cup E_{e_2}$. Obviously, {\it $\mathcal{G}_X$ is a DAG}.
We can also define $\mathcal{G}$ as the union of all IsA taxonomy $G_e$ for each entity $e$.

\begin{definition}[IsA taxonomy graph for cluster $X$]
The IsA taxonomy graph for cluster $X$, $\mathcal{G}_X$, is the union of all IsA taxonomy graph $G_e$ for each $e\in X$.
\end{definition}

\begin{prop}
For a set of entities $X$, $\mathcal{G}_X$ is a directed acrylic graph.
\end{prop}

\paragraph*{Problem Model}
Given $\mathcal{G}_X$, finding a best cluster label for $X$ thus is reduced to the problem of finding a least common ancestor of $X$ from $\mathcal{G}_X$. Given two nodes $u,v$ in $G$, if $u$ has a path to $v$, then $v$ is ancestor of $u$. For a set of entities $X$, a LCA in $\mathcal{G}_X$ is an ancestor of all entities in $X$ which has no descendant that is an ancestor of entities in $X$.


The direct LCA model clearly can ensure we find a general enough concept to cover all entities. However, the model may sacrifice the informativeness.
Hence, we need a more flexible model allowing us to control the tradeoff between informativeness and completeness. We introduce a coverage restraint $\zeta$ into LCA to tune the tradeoff between coverage and informativeness. Note that there may exist more than one LCA. We use $idf$ function (defined in Eq.~\ref{eq:cate_idf}.) to help select the best LCA. We propose {\it maximal $\zeta$-LCA} to reflect all these requisites.

\begin{pdef}[Maximall $\zeta$-LCA]
Given an IsA taxonomy graph $\mathcal{G}_X$ for the entity cluster $X$, find an node $a$ from $\mathcal{G}_X$ such that $a$ is the LCA of at least $\zeta|X|$ entities in $X$ and $idf(a)$ is maximized.
\end{pdef}

\paragraph*{Solution}
To find the best solution, we first give the monotonicity property of the $idf$ function defined in Eq.~\ref{eq:cate_idf}. The lemma~\ref{lem:cate_mon} states that if a category $c_1$ is ancestor of $c_2$ in $\mathcal{G}$, then $idf(c_1)\leq idf(c_2)$. It is obviously true. Because according to Eq.~\ref{eq:cate_idf}, the number of descendants of $c_1$ is no less than that of $c_2$.
The lemma suggests that bottom up level wise search solution for the maximal $\zeta$-LCA of $X$. Because the lower level (close to the entities) $\zeta-$LCA will certainly have a larger $idf$ value than the upper level.

For example, an IsA taxonomy graph $\mathcal{G}$, shows in Figure~\ref{fig:cate_graph}, compose of 4 entities and 4 categories. $c2$ is parent category of $c1$, so $c2$ is ancestor of $e1,e2,e3$. Thus $c1$ occurs in feature of $e1,e2$ and $c2$ occurs in feature of $e1,e2,e3$, then $idf(e1)=log(4/2)$, and $idf(c2)=log(4/3)$. Similarly, $c4$ is ancestor of both 4 entities, then $idf(c4)=log(4/4)$. Clearly idf of a category is always no larger than its descendant category.

Specifically, we use $L_i$ ($i\geq 1$) to denote the categories to be tested in the $i$-th level.
$L_1$ is defined as the parents of $X$ in $\mathcal{G}_X$.
In the $i-$th level, we first let $L_i$ be the parents of categories of $L_{i-1}$.
Then, we calculate the coverage of each category in $L_{i}$. If any category cover at least $\zeta|X|$ entities, we return the one with maximal $idf$ value from $L_i$ as the result. Otherwise, the procedure proceeds into the $(i+1)-$th level.
Note that in each level, we use the $idf$ function to select the most specific one among all $\zeta-$ LCA discovered in the same level. We also highlight that $L_i$ many overlap with $L_{i-1}$. The above level-wise search can certainly find the optimal solution due to Lemma~\ref{lem:cate_mon},

\begin{lemma}[monotonicity]
\label{lem:cate_mon}
Given two categories $c_1$, $c_2$, if $c_1$ is an ancestor of $c_2$ in $\mathcal{G}$, we have $idf(c_1)\leq idf(c_2)$.
\end{lemma}

\begin{example}
We give the example to show how maximal $\zeta$-LCA can be found. Suppose there is cluster $X$ compose of $e1,e2,e3$ in Figure~\ref{fig:cate_graph} and we set $\zeta=1$. First, for categories in $L_1=\{c1,c2\}$, their coverage is $0.67,0.33$ respectively. Both coverage is lower than $\zeta$, so we continue search upper level $L_2=\{c2,c4\}$, here both coverage of $c2$ and $c4$ is 1. Hence $c2$ and $c4$ satisfy the requirement of $\zeta$-LCA, we select the most specific one $c2$ as the maximal $\zeta$-LCA since idf weight of $c2$ is larger than $c4$.
\end{example}

\paragraph*{Implementation Optimizations}
In real implementations, we have two issues to address. First, we set a maximal layer limit to boost the search procedure. Second, we need to handle cases where no appropriate a $\zeta$-LCA is found. Next, we elaborate our solutions to each issue.

We set a upper limit for the search level due to two reasons. On one hand, $X$ may have no valid $\zeta$-LCA. On the other hand, even if we find a $\zeta$-LCA in a higher layer. The category we found may be too general thus is meaningless.

Note that our algorithm may return no result due to two reasons. First, the constraint posed by $\zeta$ is too stricter. Second, the upper-limit may although boosted the search but may miss some valid solution occurring in upper level. To solve this problem, we run the maximal $\zeta$-LCA search iteratively with $\zeta$ varying from $1$ to $\frac{1}{|X|}$ (with increment as $\frac{1}{|X|}$). Obviously, the iterative search can certainly find a solution if at least category occur in $\mathcal{G}_X$.


\section{Experiment}
In this section, we present our experimental results.
We run the experiments on Wikipedia (released in January 1, 2013). The basic statistics of Wikipedia before and after revoking the noisy entities are shown in Table \ref{tab:wiki_statistic}. We refer to the linked entity with at least one category as {\it valid} linked entity because we need to use the category information for the clustering. We run all experiments on a 64 bit Windows Server 2008 system with Intel Xeon E5620 @ 2.40GHz 16 cores cpu and 48G memory. We implement all the programs in Java.

We totally find 9.8M clusters for 1.95M articles. For each article, we find 5 cluster on average. Each cluster contains 3.3 entities on average. If we treat the <article entity, property, an entity in a cluster> as a single fact, we extracted overall 32M facts.



\begin{table}[h]
\centering
\small
\caption{Statistics of Wikipedia before/after removing the noisy linked entities}
\label{tab:wiki_statistic}
\begin{tabular}
{l  p{1.8cm}  p{1.8cm}}
\hline
\textbf{Item} & \textbf{before} & \textbf{after}\\
\hline
{\#article} & {3.04M} & {3.04M}\\
\hline
{\#categories} & {0.84M} & {0.84M}\\
\hline
{\#article has linked entity} & {3.01M} & {2.01M}\\
\hline
{\#linked entity per article} & {30} & {20}\\
\hline
{\#article has valid entity} & {2.21M} & {1.95M}\\
\hline
\end{tabular}
\end{table}

\subsection{Effectiveness}
In this subsection, we justify the effectiveness of our system with the comparison to two state-of-the-art systems to extract knowledge from Wikipedia. Both of the two competitors extract the relationship of entity pairs by handling natural language sentences. The first system (S1) finds the sentences in an article mentioning two entities. The sentences will be parsed to drive a dependency tree, and the shortest dependency path from one entity to the other entity gives the syntactic structure expressing the relationship between the entity pair~\cite{denpath}. However, an entity may be expressed in different formats (known as the coreference resolution problem), which results into the low recall of S1. To solve the coreference resolution problem, in the second system (S2) we borrow the idea from~\cite{denpatternpath} to extract many syntactic patterns of an entity, then use S1 to extract facts from Wikipedia.

We evaluate the precision and user satisfactory for all the systems. We randomly select 10 Wikipedia articles and recruited 5 volunteers to manually evaluate the quality of the extracted facts of these articles. We present the existing infobox as reference to them and ask them to evaluate the systems. Each volunteer was asked to rate the knowledge by one of the options in {\it perfectly sensible, well sensible, somewhat sensible, not sensible at all}. We assign each option with a score from 0({\it not sensible at all}) to 3({\it perfect sensible}).

The comparison results are shown in Table~\ref{tab:comparison}, where {\it Time cost per fact} is the average time cost on generating one fact (the preprocessing time including finding the sentences is not considered in S1 and S2). {\it Precision} is measured as the percentage of sensible knowledge (all three options except {\it not sensible at all}). {\it Recall} is the percentage of linked entities that can be found a relationship between it and the article entity.
{\it user satisfactory} is the average score for all samples.
Note that we also give the user satisfactory for the existing infobox.
\begin{table}[h]\small
\centering
\caption{Comparison to baseline systems}
\label{tab:comparison}
\begin{tabular}
{l l l l l}
\hline
\textbf{Matric} & \textbf{S1} & \textbf{S2} & \textbf{C\&L} & \textbf{Infobox}\\
\hline
{Time cost per fact(ms)} & {648.47} &{648.47} & {9.02} &{--} \\
\hline
{Precision} & {0.57} & {0.51} & {0.82} &{--} \\
\hline
{Recall} & {0.19} & {0.29} & {0.68} &{--} \\
\hline
{User Satisfactory} & {1.19} & {1.04} & {2.01} &{3} \\
\hline
\end{tabular}
\end{table}

We can see from Tbale~\ref{tab:comparison} that our system (C\&L) is significantly more efficient than the two competitor systems. Besides this, our system outperforms the competitors significantly in precision, recall and user satisfactory.
We highlight that the precision of our system is almost 0.91.
The recall of our system is 0.68. The reason is that some linked entities are regarded as noises or do not have category information and consequently can not be clustered.
If we didn't count them in the recall computation, we will get an even better recall.
The user satisfactory of our system is close to that on the existing infobox, suggesting that our extraction system has close quality to existing infobox.
Comparing to S1, S2 has a higher recall but a lower precision because it can discover more sentences containing the article entity and linked entity.

\subsection{Remove Noisy Linked Entities}
In this subsection, we evaluate the effectiveness of our rank aggregation approach.
The statics of Wikipedia after removing all unrelated linked entities are shown in Table~\ref{tab:wiki_statistic}.
To quantify the goodness of a ranking scoring, we first manually label each linked entity as {\it related} or {\it unrelated}. This manually labeled data set is used as the ground truth. Then for each ranking measure, we generate an ordering by the measures and evaluate the ordering with the comparison to the ground truth by $M@K$.
$$M@K=\frac{|M\bigcap K|}{|M|}$$
where $M$ is the set of linked entities labeled with {\it related}, and $K$ is the set of top-$K$ entities in the ordering.
By varying $K$ from 0 to the number of elements to be ordered, we can draw the curve of $M@K$.
We can further quantify the {\it closeness} of a ranking measure $r$ with respect to a range $[s, t]$as
\begin{equation}
\label{eq:closeness}
closeness(r,s,t)=\frac{\sum_{K=s}^{t}{\frac{M@K_{r}}{M@K_{truth}}}}{t-s+1}
\end{equation}
where $M@K_{r}$ and $M@K_{truth}$ are the $M@K$ curve of the measure $r$ and the ground truth, respectively, $[s, t]$ means the range from top-$s$ to top-$t$. The $closeness(r,s,t)$ actually characterizes the average closeness in the range of $[s, t]$.
When $s=1$ and $t=n$ ($n$ is the number of all elements) we have $closeness(r)$
measures the entire closeness to the ground truth of the ranking measure $r$.

\begin{figure} \centering
\subfigure[\small{Steve Jobs}] { \label{fig:steve_aggregation}
\includegraphics[width=0.45\columnwidth]{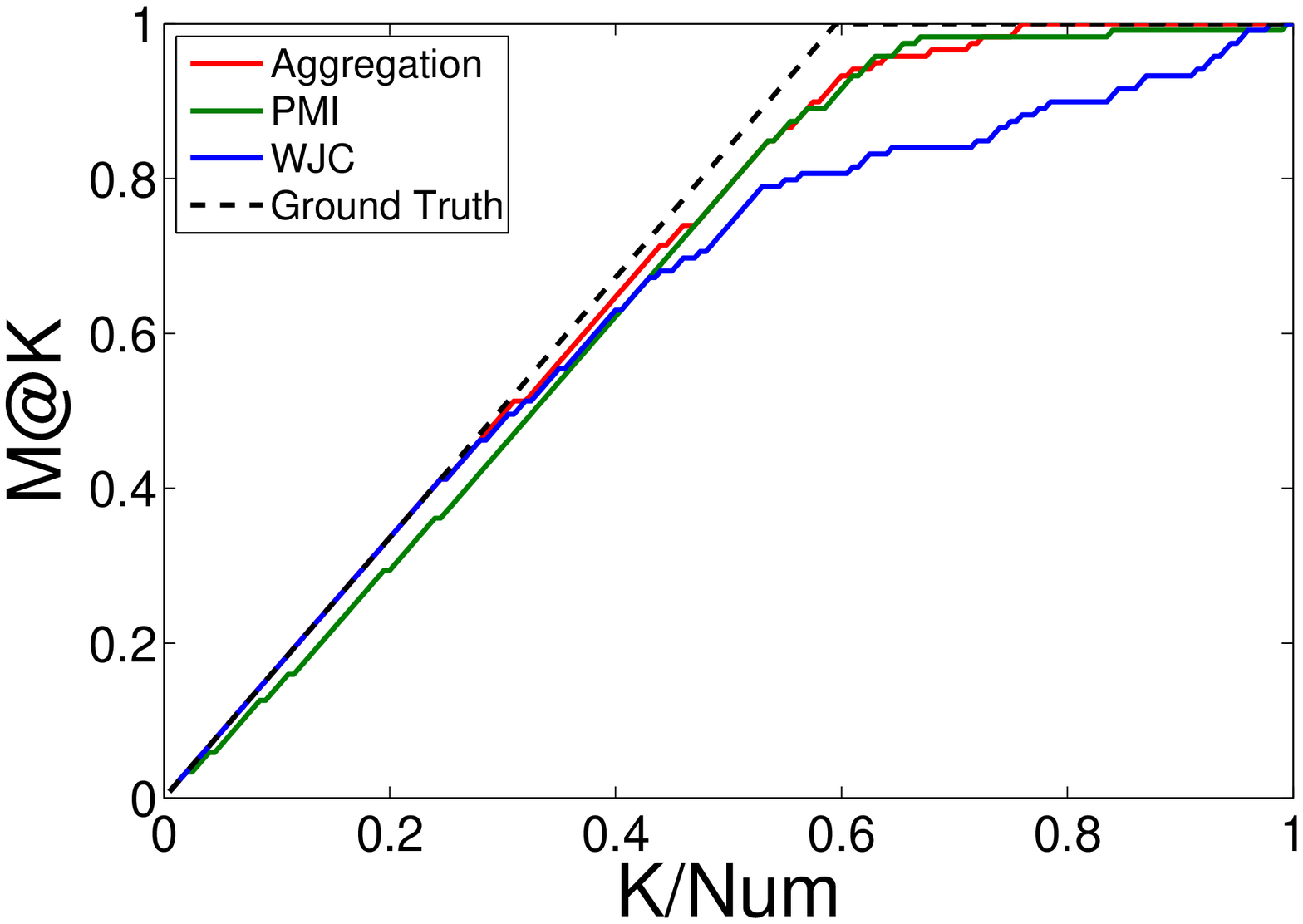}
}
\subfigure[\small{Apple Inc.}] { \label{fig:apple_aggregation}
\includegraphics[width=0.45\columnwidth]{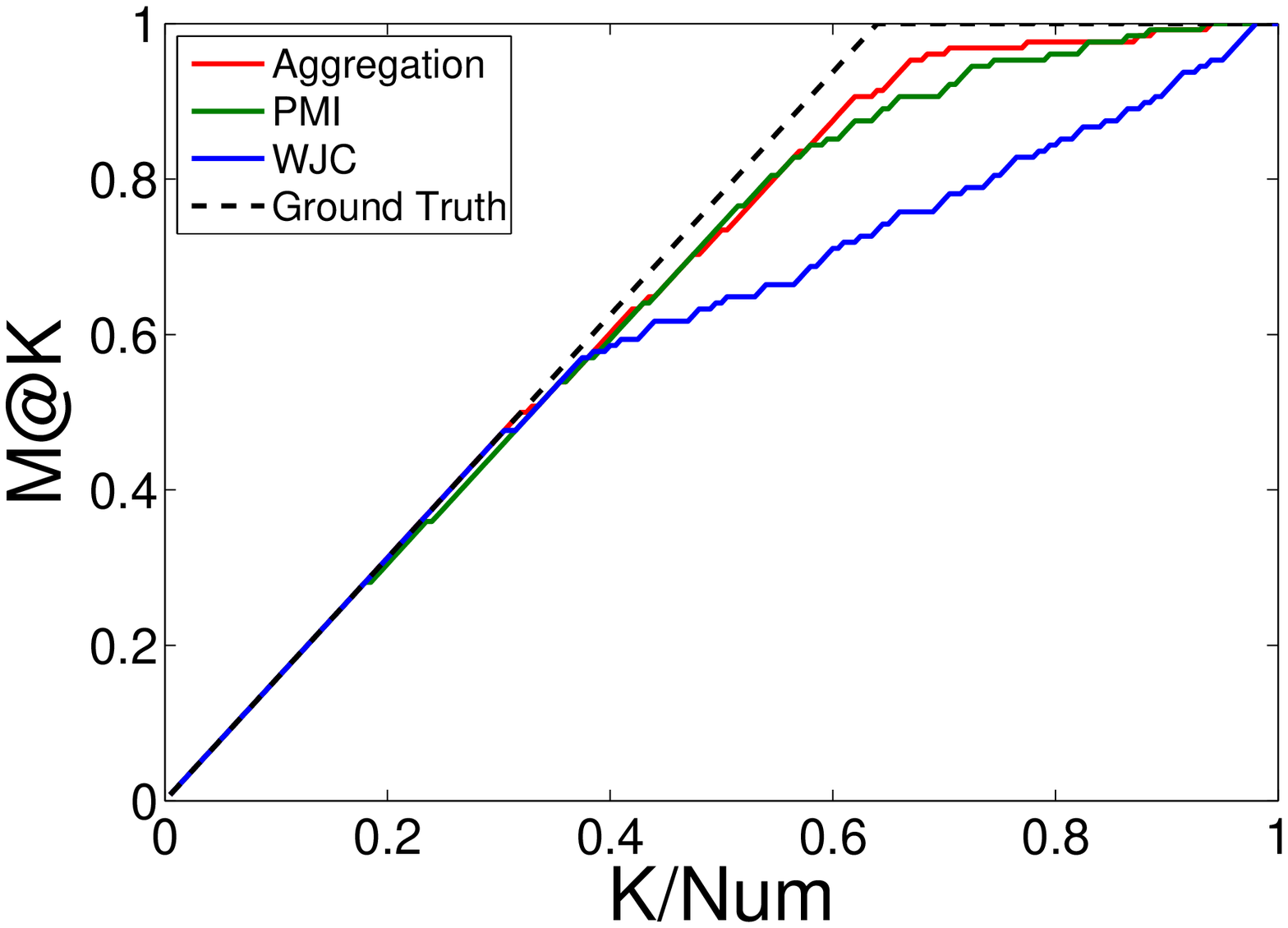}
}
\caption{\small{M@K for different ranking strategies, {\it Num} is number of entities in the ordering.}}
\label{fig:rank_aggregation_result}
\end{figure}

\paragraph*{Comparison to Individual Rankings}
We use {\it Steve Jobs, Apple Inc.} to evaluate the effectiveness of different ranking measures. Results on other articles are similar to them. In our experiment, we order linked entities of the two samples by different rankings. The $M@K$ curves are shown in Figure~\ref{fig:rank_aggregation_result}, in which we compare our aggregated measure to the two individual ranking measures: $PMI$ and $WJC$. We also give the $M@K$ curve for the ground truth. The closer to the ground truth curve the better the measure is. We can see that $PMI$ is better than $WJC$ in noise detection since $PMI$ in general is closer than $WJC$ to the ground truth curve. 
In general, the curve of our aggregated measure is closer to the ground truth curve than the two individual measures. Hence, our rank aggregation is better than either $PMI$ or $WJC$ and outperforms them in both detecting strongly related entities and recognizing noisy entities.

\begin{figure} \centering
\subfigure[\small{Steve Jobs}] { \label{fig:steve_alpha}
\includegraphics[width=0.45\columnwidth]{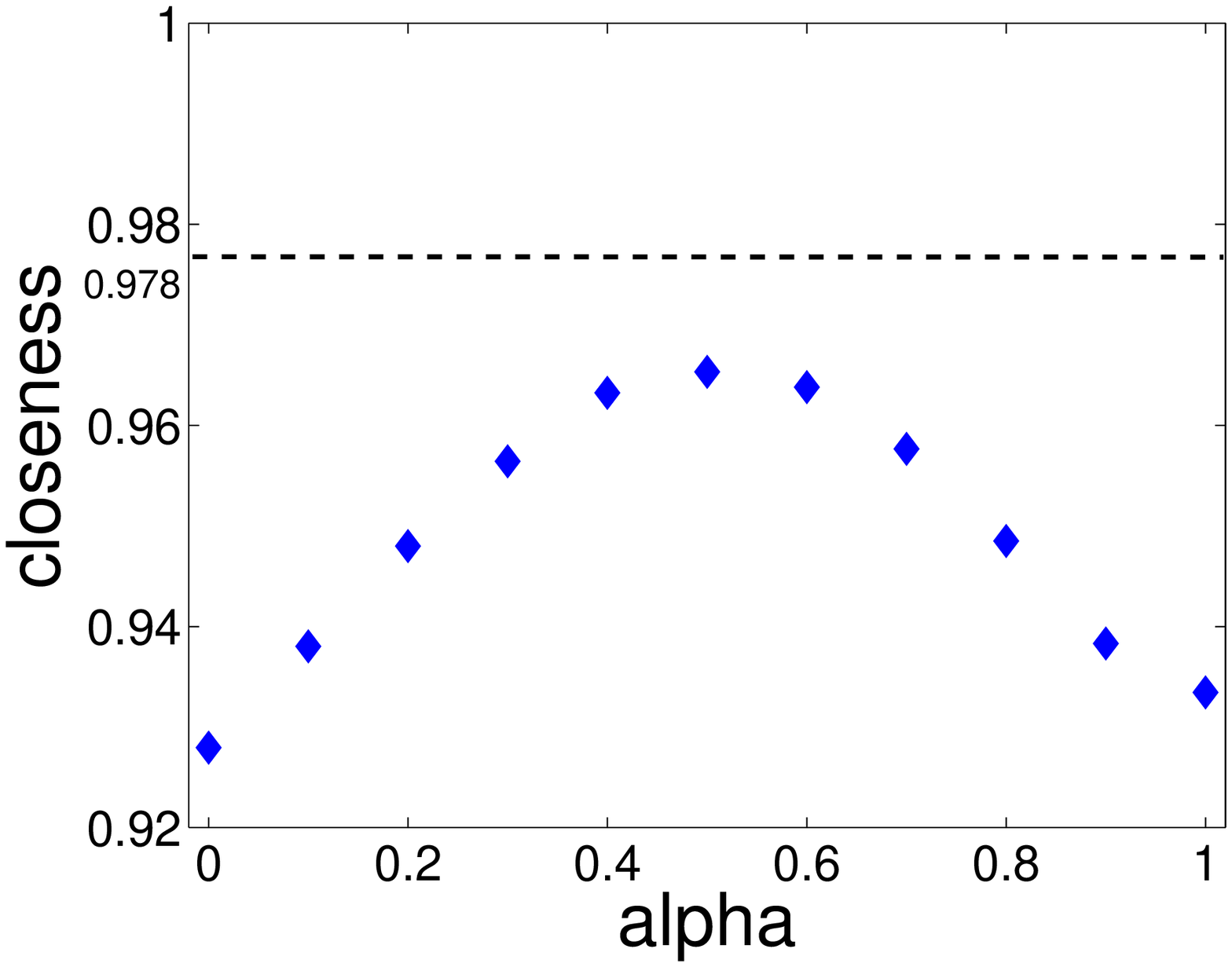}
}
\subfigure[\small{Apple Inc.}] { \label{fig:apple_alpha}
\includegraphics[width=0.45\columnwidth]{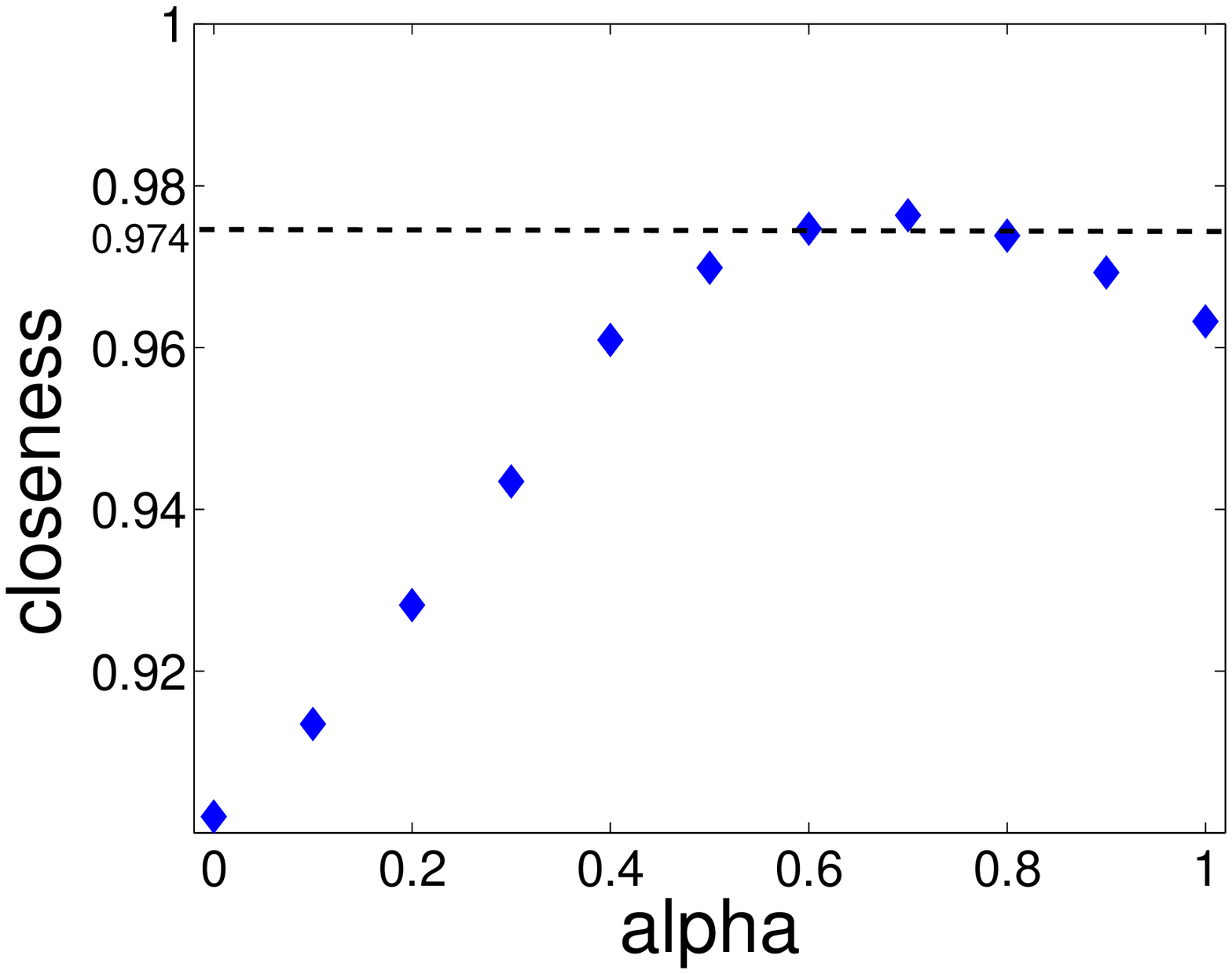}
}
\caption{\small{Comparison to other aggregated measures.}}
\label{fig:aggregation_compare}
\end{figure}

\paragraph*{Comparison to Other Aggregated Measures}
We next compare our aggregated ranking to the naive linear combination method with static $\alpha$. We vary $\alpha$ from 0 to 1 with increment of 0.1 so that we can compare to the different linearly combined measures.
For the two samples we calculate the closeness ($closeness(r)$) between the ground truth and different ordering measure $r$. The results are shown in Figure~\ref{fig:aggregation_compare}, where the horizontal line is our aggregated measure. We can see that that our aggregated is superior to the naive linearly combined measure consistently over different $\alpha$. Only in the case of {\it Apple Inc.} with $\alpha$ raining from $0.6$ to $0.8$, the linearly combined measure can reach the same goodness as our measure. But in general, users have no prior knowledge to set an appropriate value for $\alpha$. Instead our method automatically computes the appropriate $\alpha$ and achieves the best performance.

\begin{figure} \centering
\label{fig:top_last_20}
\includegraphics[width=0.6\columnwidth]{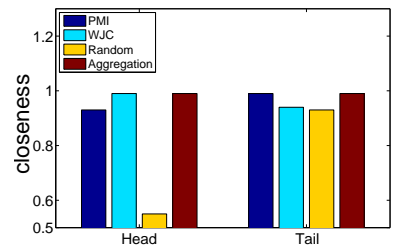}
\caption{\small{Closeness for head and tail part in the order.}}
\end{figure}

\paragraph*{Rationality of the motivation}
Next, we justify the motivation of renaming aggregation method. Recall that our aggregation is based on the fact that $PMI$ is good at identifying the semantically unrelated entities and $WJC$ is good at identifying the semantically related entities.
To verify this, we need to analyze the entities in the head and tail part of the orderings.
We select 100 articles randomly and manually label their linked entities as {\it related} and {\it unrelated}. 
For each measure, we calculate the closeness for the top 20 (head) and last 20 (tail) entities respectively by Eq.~\ref{eq:closeness}.  For comparison, we also give the result of a random ordering. The results are shown in Figure~\ref{fig:top_last_20}.
We can see that in the head part $WJC$ is better than $PMI$ and both outperforms the random ordering. But in the tail part $PMI$ is better than $WJC$ and random order.
In both head and tail part, the aggregated measure perfumes the best, which justify again the effectiveness of our ranking aggregation approach.

\subsection{Clustering and Labeling}

We first give the metrics used for the evaluation, then present the experiment results, some clustering and labeling results are shown in Table ~\ref{tab:ExtractedKnowledge}.

\paragraph*{Metrics for the Evaluation of Clustering}
To evaluate the effectiveness of a cluster, we use both the subjective and objective metric. The objective metrics include the {\it inter-cluster distance} (average distance between cluster centers) and {\it intra-cluster distance} (average distance between entities and corresponding cluster center). The two individual metrics can be furthered combined as a synthesis score, known as {\it valid} index.  Formally, let $K$ be the number of clusters, $m_i$ be the center of cluster $C_i$, we have
\begin{equation}
inter = \frac{2}{K(K-1)}\sum\limits_{i=1}^{K}\sum\limits_{j=i+1}^{K}dis(m_i,m_j)
\end{equation}
\begin{equation}
intra = \frac{1}{K}\sum\limits_{i=1}^{K}\frac{1}{|c_j|}\sum\limits_{e\in c_j}dis(m_j,e)
\end{equation}
\begin{equation}
\label{eq:valid}
valid = \frac{inter}{intra}
\end{equation}
 A good clustering result has a large inter distance and a small intra distance, which induces a large valid index.

When the cluster is labeled, we may alternatively use subjective metric to evaluate the quality of the clustering. We adopt {\it precision} to evaluate the quality of the extracted knowledge. For a certain entity, suppose its linked entities are clustered into $C=\{C_1,...C_k\}$ and each cluster $C_i$ has label $l_i$. The $precision$ of $C$ under label set $L=\{l_i\}$ is defined as:
\begin{equation}
\label{eq:pcl}
P(C,L)=\frac{1}{k}\sum_{C_i\in C}{\frac{match(C_i,l_i)}{|C_i|}}
\end{equation}
Where $match(C_i,l_i)$ is the percentage of entities in cluster $C_i$ that can be appropriately labeled by the $l_i$. $match(C_i,l_i)$ is evaluated by humans.

\paragraph*{Metric for the Labeling Evaluation}
Given a cluster $C=\{C_i\}$ and their label set $L=\{l_i\}$, we use the following metrics to evaluate the accuracy of $L$ with respect to $C$.
\begin{itemize}
\item {\it Coverage}. Coverage of $l_i$ with respect to $C_i$ is the percentage of entities in $C_i$ which is the descendant of $l_i$ in the IsA taxonomy graph {\it $\mathcal{G}_c$}. Thus, the coverage of $L$ with respect to $C$ is the average coverage of each label $l_i$ with respect to corresponding $C_i$.
\item {\it Correctness}. We use $P(C,L)$ to measure correctness of $L$ with respective to $C$.
\end{itemize}

\paragraph*{Clustering Results}
To evaluate the performance of clustering, we cluster the linked entities for {\it China, Shanghai, Apple Inc., Steve Jobs, Barack Obama, New York City}, and using our clustering approach with $\alpha=0.0001$ and $iteration=5$. We give the results in Table~\ref{tab:cluster_evaluate}. To calculate $P(C, L)$, we use the labels generated by {\it maximal $\zeta$-LCA}. 
We can see that average valid of clusters is around 2.0, and average precision is approach to 90\%, which suggests that the generated clusters are of high quality.

\begin{table}[h]\small
\centering
\caption{Evaluation of clustering results}
\label{tab:cluster_evaluate}
\begin{tabular}
{p{2cm} p{1cm} p{1cm} p{1cm} p{0.6cm} p{1cm} }
\hline
\textbf{Entity} & \textbf{\#linked Entity} & \textbf{\#cluster} & \textbf{Time (ms)} & \textbf{Valid} & \textbf{P(C,L)}\\
\hline
{Shanghai} & {276} & {36} & {1333} & {2.88} & {0.94}\\
\hline
{Steve Jobs} & {298} & {35} & {2047} & {1.95} & {0.79}\\
\hline
{Apple Inc.} & {315} & {34} & {1533} & {2.30} & {0.88}\\
\hline
{Barack Obama} & {395} & {49} & {4359} & {2.01} & {0.96}\\
\hline
{China} & {508} & {67} & {5850} & {2.20} & {0.95}\\
\hline
{New York City} & {586} & {78} & {9330} & {2.06} & {0.87}\\
\hline
{Average} & {396} & {49} & {4075} & {2.23} & {0.89}\\
\hline
\end{tabular}
\end{table}

\paragraph*{Labeling Results}
We compare our labeling approaches to the baseline approaches including:  {\it MF, MFI} and a state-of-the-art approach {\it Score Propagation(SP)}~\cite{Cluster_Label}. SP uses Wikipedia as external source from which candidate cluster labels can be extracted. Given a cluster, SP first generate some concepts and categories as candidate labels from Wikipedia by measuring the relevance to terms in the cluster. 
For a cluster, SP first calculate the frequency score of keywords in all candidate labels, then propagate the score from keywords to label. Finally the label with highest score is selected as the cluster label.

\begin{figure} \centering
\subfigure[\small{{\it Coverage} for different labeling strategies}] { \label{fig:coverage_metric}
\includegraphics[width=1.1\columnwidth]{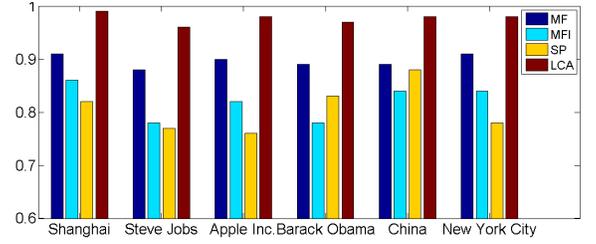}
}
\subfigure[\small{{\it Correctness} for different labeling strategies}] { \label{fig:correct_metric}
\includegraphics[width=1.1\columnwidth]{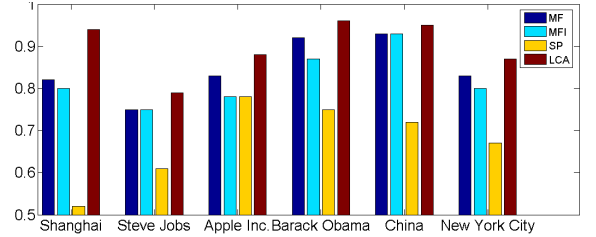}
}
\caption{Evaluation of cluster labeling strategies.}
\label{fig:labeling_evaluate}
\end{figure}

We run {\it maximal $\zeta$-LCA} with $\zeta=0.8$. For clusters generated from linked entities of above 6 sample entities, we use {\it coverage} and {\it correctness} to evaluate different labeling strategies. The results are shown in Figure~\ref{fig:labeling_evaluate}.

We can see from the Figure~\ref{fig:labeling_evaluate} that coverage of $MF$ is larger than $MFI$ and $SP$, that is reasonable because the category voted by $MF$ is the feature of most entities in the cluster. And {\it maximal $\zeta$-LCA} has the largest coverage which approach to 100\%, because the selected category is at least the ancestor of 80\% entities in the cluster. For correctness, $MF$ is a little better than $MFI$, and obviously outperform $SP$, and also {\it maximal $\zeta$-LCA} performs better than other approaches.

\begin{table}[h]\tiny
\centering
\caption{\small{Labeled clusters generated from {\it Apple Inc.}, line in column {\it Label} represents {\it MF, MFI, SP and Maximal $\zeta$-LCA} separately, and each label is given with its {\it Coverage}}}
\label{tab:apple_inc_cluster}
\begin{tabular}{|p{0.2cm} | l | l |}
                \hline
                    \textbf{No.} & \textbf{Cluster} &\textbf{Label}\\
                \hline
                    \multirow{5}{*}{1} & Alan Kay & people by status (1.0)\\
                    & Gil Amelio & tunisian-jewish descent (0.2)\\
                    & Andy Hertzfeld & people(1.0)\\
                    & Ronald Wayne  & \bf{apple inc. employees (0.8)}\\
                    & Guy Kawasaki & \\
                \hline
                    \multirow{4}{*}{2} & 3G, BBC Online & tele conferencing (0.5)\\
                     & Electronic product environ- & tele conferencing (0.5)\\
                     & mental assessment tool & open standards (0.25)\\
                     & Enhanced data rates- & \bf{electricity(1.0)}\\
                     & for gsm evolution & \\
                \hline
                    \multirow{4}{*}{3} & Google Maps & ios software (1.0)\\
                    & ios 6 & ios software (1.0)\\
                    & iBooks & ios software (1.0)\\
                    & xSan, iTunes & \bf{ios software (1.0)}\\
                \hline
                    \multirow{4}{*}{4} & Dell & computer hardware companies (1.0) \\
                    & Foxconn & computer hardware companies (1.0) \\
                    & IBM & computer hardware companies (1.0) \\
                    & Intel & \bf{computer hardware companies (1.0)} \\
                \hline
\end{tabular}
\end{table}

We also give the clustering results for {\it Apple Inc.} under different labeling approaches in Table~\ref{tab:apple_inc_cluster}. We can see that MI in general can find the frequent but general category, such as the first cluster. MFI tends to find the specific label which in general has a low coverage, such as the second cluster. The performance of SP is not stable, which may generate either the general or specific label (see the first and second clusters of SP). Compared to these methods, {\it maximal $\zeta$-LCA} method can generate specific label of high coverage in most clusters.
{\it Maximal $\zeta$-LCA} enables us to find knowledge such as {\it <apple inc., ios software, \{google maps, ios 6, ibooks, xsan, itunes\}>}.


\section{Related works}

\nop{
\remind{Kezun: could you double check all the reference. some authors names look very strange. some are wrong, such as ref (1) }
\remind{we also need to survey: (a) data mining on encyclopedia in general; (b) IE in general; and (c) document summarization}
}

\paragraph*{Data mining on encyclopedia}
Many works have been done in online encyclopedia to achieve some applications, especially in Wikipedia, one of the most valuable online data source. ESA\cite{ESA} and WikiRelate\cite{WikiRelate} use Wikipedia to compute semantic relatedness for an entity pair. And \cite{CatDoClustering} and \cite{Cluster_Label} use Wikipedia as external knowledge for clustering or labeling cluster, which enrich the representation of document with additional features from Wikipedia.

\paragraph*{Structural knowledge extraction}
In the work of structural knowledge extraction. KnowItAll~\cite{KnowItAll} and Textrunner~\cite{Textrunner} extract open information from free text, and some challenging task such as NER, dependency parsing, and relationship extraction are commonly use in text analysis. Some structural knowledge have also been extracted from Wikipedia, like YAGO~\cite{YAGO} and DBpedia~\cite{dbpedia}. DBpedia represents in RDF, is a large scale structured knowledge base, who extracts structured information from Wikipedia, and also links to other datasets on the Web to Wikipedia. But DBpedia is built on existing infobox in Wikipedia and structural knowledge in other datasets. To make Wikipedia more structural, Semantic Wikipedia~\cite{semantic_wikipedia} proposes a formalism to structure Wikipedia's content as a collection of statements, the statement can explain the relationship between article and linked entities. And~\cite{denpatternpath} try to extract relationship of linked entity use syntactic and semantic information, and refer to relationships in infobox. These article-related relationships can be good complement for infobox. Specifically, to supply attribute value for incomplete infobox, Kylin~\cite{Kylin}, iPupulator~\cite{iPopulator} and IBminer~\cite{IBminer} learn models from structured information to guide the text processing. For example, Kylin first predicts what attributes a sentence may contain, and further use CRF to extract attribute values from the candidate sentences.

\paragraph*{Document summarization}
Instead of mining relationship of single linked entity, we focus on all the linked entities for an article. Since each linked entity direct to a specific article in Wikipedia, multi-document summarization is a good solution to handle it. We can summarize the linked entities to groups and generate a theme for each group. In document summarization, ~\cite{doc_sum1} selects important sentences or paragraphs in the set of documents and build a summary with these passages. And ~\cite{LDA_summarize} forms the summary of documents to different event theme by using LDA to capture the events being covered by the documents. Clustering is another widely used method to do summarization, such as XDoX~\cite{XDoX}, and select a representative passage from the cluster after clustering.

In this paper, we use clustering method to summarize linked entities. And different from above structural knowledge extraction methods, we use the structured information(linked entities, and categories) only in Wikipedia to extract knowledge(infobox). In this way, we can avoid the text processing problem such as NER and dependency parsing. 
\section{Conclusion}
Discovering and enriching structural information in online encyclopedia is valuable and challenging work. Different from previous free-text focused methods, in this paper, we propose an novel, semi-structured information based approach. We extract knowledge from Wikipedia using rich set of linked entities.

We propose an {\it cluster-then-label} approach, which clusters the linked entities into different semantic groups, and then give each group a semantic label (a property). In this way, we can get groups of facts in the form of cluster and semantic label. We further propose a novel position aware rank aggregation method to detect the semantic related entities. We also propose an effective cluster reuse strategy to run clustering for millions of entities in Wilkipeida. With these effective and efficient approach, we extracted 18 million new facts from Wikipedia.

\bibliographystyle{abbrv}
\bibliography{refer}

\end{document}